\documentclass[a4paper,twocolumn,11pt,unpublished]{quantumarticle}
\pdfoutput=1
\usepackage[utf8]{inputenc}
\usepackage[english]{babel}
\usepackage[T1]{fontenc}
\usepackage{amsmath,amssymb}
\usepackage[breaklinks]{hyperref}
\usepackage{tikz}
\usepackage{lipsum}
\usepackage[numbers,sort&compress]{natbib}
\usepackage[normalem]{ulem}
\usetikzlibrary{automata}
\usetikzlibrary{quantikz}
\usetikzlibrary{trees}

\begin{document}

\title{A quantum unstructured search algorithm for discrete optimisation: the
  use case of portfolio optimisation}

\author{Titos Matsakos}
\affiliation{Financial Risk Analytics, Market Intelligence, S\&P Global,
  25 Ropemaker St, London, EC2Y\,9LY, UK}
\orcid{0000-0003-0447-2147}
\email{titos.matsakos@spglobal.com}
\author{Adrian Lomas}
\affiliation{Financial Risk Analytics, Market Intelligence, S\&P Global,
  25 Ropemaker St, London, EC2Y\,9LY, UK}
\email{adrian.lomas@spglobal.com}

\maketitle

\begin{abstract}
We propose a quantum unstructured search algorithm to find the extrema or roots
of discrete functions, $f(\mathbf{x})$, such as the objective functions in
combinatorial and other discrete optimisation problems.
The first step of the Quantum Search for Extrema and Roots Algorithm (QSERA) is
to translate conditions of the form $f(\mathbf{x}_*) \simeq f_*$, where $f_*$ is
the extremum or zero, to an unstructured search problem for $\mathbf{x}_*$.
This is achieved by mapping $f(\mathbf{x})$ to a function $u(z)$ to create a
quantum oracle, such that $u(z_*) = 1$ and $u(z \neq z_*) = 0$.
The next step is to employ Grover's algorithm to find $z_*$, which offers a
quadratic speed-up over classical algorithms.
The number of operations needed to map $f(\mathbf{x})$ to $u(z)$ determines the
accuracy of the result and the circuit depth.
We describe the implementation of QSERA by assembling a quantum circuit for
portfolio optimisation, which can be formulated as a combinatorial problem.
QSERA can handle objective functions with higher order terms than the
commonly-used Quadratic Unconstrained Binary Optimisation (QUBO) framework.
Moreover, while QSERA requires some a priori knowledge of the extrema of
$f(\mathbf{x})$, it can still find approximate solutions even if the conditions
are not exactly satisfied.
\end{abstract}

\section{Introduction}

\subsection{Quantum optimisation}

Quantum computing has several applications in finance \cite[for reviews, see][]
{Orus+2019, Egger+2020b, Albareti+2022, Gomez+2022, Herman+2022, Dalzell+2023,
Herman+2023a, Intallura+2023, WilkensMoorhouse2023}.
Two of the most promising areas are Monte Carlo simulations (e.g. derivative
pricing \cite{Rebentrost+2018, Stamatopoulos+2020, Tang+2020,
CarreraVazquezWoerner2021, Chakrabarti+2021, Doriguello+2021, HanRebentrost2022}
and risk management \cite{WoernerEgger2019, Egger+2020a, Kaneko+2021,
Alcazar+2022, Stamatopoulos+2022, MatsakosNield2024}) and optimisation problems
(e.g. portfolio optimisation \cite{RebentrostLloyd2018,
VenturelliKondratyev2018, Kerenidis+2019, Hodson+2019,
PhillipsonSinghBhatia2020, Gilliam+2021, Palmer+2021, Slate+2021,
BakerRadha2022, Mugel+2022, Brandhofer+2023, Herman+2023b, Yilmaz+2023, 
LimRebentrost2024} and transaction settlement optimisation \cite{Braine+2021}).

Optimisation problems are prevalent across scientific fields and they are also
at the core of most machine learning models.
When the objective function depends on continuous variables, gradient descent
algorithms are among the commonly-used approaches.
Two of the challenges using such algorithms are the presence of constraints and
the convergence to a local extremum rather than the global one.
These algorithms, however, are not applicable when the variables are binary and
derivatives cannot be defined, as in the case of combinatorial optimisation;
these are NP-hard problems even without constraints \cite[see][for reviews]
{Herman+2022, Herman+2023a}.
Many such problems can be formulated using the Quadratic Unconstrained Binary
Optimisation (QUBO) framework or the equivalent Ising model, both of which model
the individual contribution of each variable (first order terms) as well as
their pair-wise interactions (second order terms).
Constraints can be included as two possible types: either as ``soft'', which
adds terms to the objective function but cannot guarantee they are exactly
satisfied, or as ``hard'', which represent additional (in)equalities
\cite{BakerRadha2022}.
Depending on the form of the objective function and potential constraints, there
are several families of quantum optimisation algorithms available, such as
Quantum Adiabatic Algorithms (QAA) \cite[e.g.][]{Farhi+2000}, Variational
Quantum Algorithms (VQAs) \cite[e.g.][]{Peruzzo+2014, Farhi+2014}, Quantum
Linear System Algorithms (QLSAs) \cite[e.g.][]{Harrow+2009, Dervovic+2018}, and
Quantum Search Algorithms (QSA) \cite[e.g.][]{DurrHoyer1996, Bulger+2003,
Baritompa+2005} (see \cite{Abbas+2024} for a review of quantum optimisation).

In QAA, the objective function is represented by a Hamiltonian,
$\mathcal{H}_\mathrm{c}$, and its minimum by the corresponding ground state.
The quantum computer is initialised in the known ground state of another
Hamiltonian, $\mathcal{H}_0$, which is then evolved towards
$\mathcal{H}_\mathrm{c}$ slowly enough such that the system remains at the
ground state.
Measuring the final state returns the minimum value of the objective function,
see App.~\ref{app:qaa}.

In VQAs, on the other hand, the values of the objective function are represented
by parametrised states.
Such algorithms can be implemented in hybrid systems where: i) a parametrised
state is initialised and measured in a quantum computer, ii) a classical
computer processes the output and updates the state initialisation parameters
such that the next iteration gets closer to the optimal value, and iii) these
steps are repeated until convergence.
VQAs are promising for near-term quantum applications because splitting out the
parameter optimisation steps (classical computation) from the estimation of the
objective function (quantum computation) keeps the quantum circuit depth
shallow, thus mitigating the effects of noise.
Among VQAs is the Variational Quantum Eigensolver (VQE) \cite{Peruzzo+2014},
which searches for the eigenstate that corresponds to the optimum of the
objective function (see App.~\ref{app:vqe}), and the Quantum Approximate
Optimisation Algorithm (QAOA) \cite{Farhi+2014}, whose quantum operators
resemble QAA but which are parametrised such that the ground state can be found
with a variational approach (see App.~\ref{app:qaoa} and \cite{Blekos+2024} for
a recent review).

QSAs are based on Grover's algorithm --- hereafter Quantum Unstructured Search
Algorithm (QUSA) --- which finds the desired state by flagging it with a quantum
oracle and amplifying its amplitude \cite{Grover1996}.
The corresponding optimisation algorithms, the family of Grover Adaptive Search
(GAS) algorithms, are based on the following steps: i) set a threshold value,
ii) use an oracle to identify all states less/greater than the threshold value,
iii) amplify the amplitude of these states, iv) update the threshold value from
one of them, and iv) repeat the steps until the optimum is found
\cite{DurrHoyer1996, Bulger+2003, Baritompa+2005} (see also \cite{Miyamoto+2019,
Albino+2023} for some recent extensions).
QSERA, which is described in the next section, belongs to the QSA family but is
not a GAS algorithm.

\subsection{Portfolio optimisation}

Consider a pool of $N_\mathrm{a}$ assets, with the annual expected return and
volatility of asset $i$ denoted with $\mu_i$ and $\sigma_i$, respectively.
Suppose that we have an amount $v_\mathrm{p}$ to invest to construct a portfolio
of $N_\mathrm{p} \leq N_\mathrm{a}$ assets, with $v_i$ the investment in asset
$i$.
The portfolio expected return, $\mu_\mathrm{p}$, and volatility,
$\sigma_\mathrm{p}$, are:
\begin{align}
\mu_\mathrm{p}
&= \sum_{i=0}^{N_\mathrm{a}-1}\frac{v_i}{v_\mathrm{p}}\mu_i\,,
\end{align}
\begin{align}
\sigma_\mathrm{p}^2
&= \sum_{i_1=0}^{N_\mathrm{a}-1}\sum_{i_2=0}^{N_\mathrm{a}-1}
  \frac{v_{i_1}}{v_\mathrm{p}}\frac{v_{i_2}}{v_\mathrm{p}}
  \sigma_{i_1}\sigma_{i_2}\rho_{i_1i_2}\,,
\end{align}
where $\sum_iv_i = v_\mathrm{p}$.
Limiting the pool of assets to publicly-listed equities and bonds, the number of
available assets is on the order of $N_\mathrm{a} \sim 10^5$.

Portfolio optimisation \cite{Markowitz1952} aims to find the optimal allocation, 
i.e. the values of the continuous variables $v_i$ that optimise an objective 
function under the constraint $\sum_iv_i = v_\mathrm{p}$; or, more generally,
$\sum_iv_i \leq v_\mathrm{p}$.
This is achieved either by setting a target $\mu_\mathrm{p}$ and finding the
allocation that minimises $\sigma_\mathrm{p}^2$, or by setting a target
$\sigma_\mathrm{p}^2$ --- representing the risk appetite of the investor --- and
maximising $\mu_\mathrm{p}$.
Other approaches include the maximisation of an objective function of the form:
$\lambda \mu_\mathrm{p} - (1 - \lambda)\sigma_\mathrm{p}^2$, where
$\lambda \in [0, 1]$ captures the weight the investor puts on returns versus
risk.
The problem becomes more complex when additional investment considerations are
taken into account, such as the liquidity of the assets, transactions costs,
frequency of portfolio rebalancing, and when private equity/debt assets are also
included.

The objective function can be simplified if we assume that the same amount is
invested in each asset, $v = v_\mathrm{p}/N_\mathrm{p}$, which replaces the
continuous variables $v_i$ with the discrete (binary) variables $x_i\in\{0,1\}$
representing whether an asset is selected.
In this case, the portfolio expected return and volatility are:
\begin{align}
\mu_\mathrm{p}
&= \frac{1}{N_\mathrm{p}}\sum_{i=0}^{N_\mathrm{a}-1}x_i\mu_i\,,
\label{eq:mu_p}
\end{align}
\begin{align}
\sigma_\mathrm{p}^2
&= \frac{1}{N_\mathrm{p}^2}
  \sum_{i_1=0}^{N_\mathrm{a}-1}\sum_{i_2=0}^{N_\mathrm{a}-1}
    x_{i_1}x_{i_2}\sigma_{i_1}\sigma_{i_2}\rho_{i_1i_2}\,,
\label{eq:sigma_p}
\end{align}
and the constraint is $\sum_ix_i = N_\mathrm{p}$.

To highlight how challenging even this simpler problem is, notice that we can
concatenate the values $x_i$ to form a binary string, the length of which is
$N_\mathrm{a}$.
For a pool of assets that only include publicly-listed companies, this implies
that there are $2^{10^5}$ possible investment combinations to chose from, which
is on the order of $10^{30,000}$.

The paper is structured as follows.
In Sect.~\ref{sec:qsera} we describe QSERA and how to assemble the corresponding
quantum gates for its implementation in quantum circuits.
In Sect.~\ref{sec:discussion} we discuss the advantages and limitations of the
algorithm and in Sect.~\ref{sec:application} we present an end-to-end example
based on a simple portfolio optimisation problem.
We summarise our conclusions in Sect.~\ref{sec:conclusions}.

\section{Quantum Search for Extrema and Roots Algorithm (QSERA)}
\label{sec:qsera}

\subsection{Overview of QUSA}

Suppose we want to search through an unordered list of $N$ items --- which we
enumerate with the integers $z \in \{0,1,...,N-1\}$ --- to find one specific
item, $z_*$.
To identify $z_*$, we have at our disposal a function, $u(z)$, which we can use
to check any given $z$ based on the property: $u(z_*) = 1$ and $u(z) = 0$ for
all $z \neq z_*$.
Classical computing algorithms rely on a brute-force approach: i) pick a value
$z$ from the list, ii) check it with $u(z)$, iii) if it is not $z_*$ remove it
from the list, iv) repeat until $z_*$ is found.
The probability of finding $z_*$ in the first draw is $\frac{1}{N}$, in the
second $\frac{N-1}{N}\frac{1}{N-1} = \frac{1}{N}$, in the third
$\frac{N-1}{N}\frac{N-2}{N-1}\frac{1}{N-2} = \frac{1}{N}$, etc.; i.e. it is
always $\frac{1}{N}$.
If we draw $N/2$ items from the list, the probability of finding $z_*$ is the
sum of finding it in one of these draws, i.e. $(N/2)(1/N) = 1/2$; hence, the
number of required iterations is on the order $\mathcal{O}(N)$.

In a quantum computer, $u(z)$ can be represented by a quantum oracle.
QUSA can find $z_*$ with a quadratic speed-up as compared to classical
algorithms, i.e. the number of iterations is on the order of
$\mathcal{O}(\sqrt{N})$ \cite{Grover1996}.
To show this, consider $N = 2^K$ and a quantum register of $K$ qubits,
$\ket{0}^{\otimes K}$.\footnote{
  We use the notation
  $\ket{0}^{\otimes2} = \ket{0}\otimes\ket{0} = \ket{0}\ket{0} = \ket{00}$.}
Every element of the list can be represented by a quantum state,
$\ket{z} = \ket{\mathrm{b}_{K-1}...\mathrm{b}_1\mathrm{b}_0}$,
where $\mathrm{b}_{K-1}...\mathrm{b}_1\mathrm{b}_0$ is a binary number with
$\mathrm{b}_k$ the $k$-th digit which is either 0 or 1.
Start by initialising the register in a uniform superposition of all states:
\begin{align}
\ket{+}^{\otimes K}
&= \mathcal{Q}_H\ket{0}^{\otimes K}
= \frac{1}{\sqrt{N}}\sum_{z=0}^{N-1}\ket{z}\,,
\label{eq:initial}
\end{align}
where $\mathcal{Q}_H = \bigotimes_{k=0}^{K-1}H\ket{0_k}$ and $H$ the Hadamard
gate.
Then, define a ``reflection'' unitary operator that flips the phase of the state
$\ket{z_*}$:
\begin{align}
\mathcal{Q}_*\ket{z}
  &= (-1)^{u(z)}\ket{z}
  = \big(I - 2\ket{z_*}\bra{z_*}\big)\ket{z}\,,
\label{eq:reflection}
\end{align}
and another ``reflection'' unitary operator:
\begin{align}
\mathcal{Q}_+
&= \left(2\ket{+}^{\otimes K}\bra{+}^{\otimes K} - I\right) \nonumber\\
&= \mathcal{Q}_H
  \left(2\ket{0}^{\otimes K}\bra{0}^{\otimes K} - I\right)
  \mathcal{Q}_H^\dagger \nonumber\\
&= \mathcal{Q}_H\mathcal{Q}_0\mathcal{Q}_H^\dagger\,,
\end{align}
where $\mathcal{Q}_H^\dagger = \mathcal{Q}_H$ (because $H^\dagger = H$).
The effect of the operator $\mathcal{Q}_0$ is:
\begin{align}
\mathcal{Q}_0\ket{z = 0} &= \ket{z = 0}\,, \\
\mathcal{Q}_0\ket{z \neq 0} &= -\ket{z \neq 0} \,,
\end{align}
i.e. it flips the phase of all states $\ket{z} \neq \ket{0}$.
When this pair of operators is applied repeatedly $m$ times on the initial
superposition, $(\mathcal{Q}_+\mathcal{Q}_*)^m\ket{+}^{\otimes K}$, it amplifies
the amplitude of the state $\ket{z_*}$ thus increasing the probability of
measuring it with every iteration.
The optimal number of iterations is
$m_\mathrm{opt} \simeq (\pi\sqrt{N} - 2)/4$ (see App.~\ref{app:qusa_example})
and the corresponding quantum circuit is:
\begin{center}
\begin{quantikz}[column sep=0.2cm]
\lstick{$\ket{0}^{\otimes K}$}
  & \gate{\mathcal{Q}_H}\qwbundle[alternate]{}
  & \gate{\mathcal{Q}_*}\qwbundle[alternate]{}
    \gategroup[1, steps=5, style={dashed, rounded corners, inner xsep=0,
      inner ysep=0}]{$(\mathcal{Q}_+\mathcal{Q}_*)^{m_\mathrm{opt}}$}
  & \gate{\mathcal{Q}_+}\qwbundle[alternate]{}
  & \cdots\qwbundle[alternate]{}
  & \gate{\mathcal{Q}_*}\qwbundle[alternate]{}
  & \gate{\mathcal{Q}_+}\qwbundle[alternate]{}
  & \qwbundle[alternate]{}\rstick{$\ket{z_*}$}
\end{quantikz}
\end{center}

\subsection{Description of QSERA}

Consider a discrete function of the following form:
\begin{align}
&f(\mathbf{x})
= w + \sum_{i_1}\sum_{i_2}\cdots\sum_{i_K}
  w_{i_1i_2...i_K}'x_{i_1}x_{i_2}\cdots x_{i_K} \nonumber\\
&= w + \sum_{i}w_ix_i
  + \sum_{i_1}\sum_{i_2>i_1}w_{i_1i_2}x_{i_1}x_{i_2} \label{eq:f_pairs}\\
  &\quad+ \sum_{i_1}\,\,\,\,\sum_{i_2>i_1}\sum_{i_3>i_1,i_2}\!\!
    w_{i_1i_2i_3}x_{i_1}x_{i_2}x_{i_3}
  + \cdots\nonumber\\
  &\quad+ \sum_{i_1}\sum_{i_2>i_1}\cdots
    \!\!\!\!\!\!\!\!\sum_{i_K>i_1,...,i_{K-1}}\!\!\!\!\!\!\!\!\!
    w_{i_1i_2...i_K}x_{i_1}x_{i_2}\cdots x_{i_K}\,,\nonumber
\end{align}
where $\mathbf{x}$ is a $K$-dimensional vector, $x_i\in\{0,1\}$ is its $i$-th
element, $w$ and $w_{i_1i_2...i_K}'$ are constants, and the terms
$w_{i_1i_2...i_K}'x_{i_1}x_{i_2}\cdots x_{i_K}$ can be reorganised into $2^K-1$
terms with parameters $w_i, w_{i_1i_2}, ..., w_{i_1i_2...i_K}$ with distinct
subscripts in ascending order.\footnote{
  The second equality of Eq.~(\ref{eq:f_pairs}) follows from the property
  $x_i^n = x_i$ for any positive integer $n \leq K$; e.g., the first order terms
  are  obtained when $i_1 = i_2 = \cdots = i_K$, i.e.
  $w_ix_i = w_{ii...i}'x_i^K$.}
Equation~(\ref{eq:f_pairs}) has up to $K$-th order terms and thus describes more
general objective functions than the QUBO framework, which only has first and
second order terms.
QUBO is as a special case of Eq.~(\ref{eq:f_pairs}) when the values of $w$ and
all other parameters with more than two indices are zero.
Note that the variable $\mathbf{x}$ can be expressed as a binary number,
$\mathrm{b}_{K-1}...\mathrm{b}_1\mathrm{b}_0$ with $\mathrm{b}_k$ taking the
value of $x_k$, and hence can be represented by an integer $z$, i.e.
$\mathbf{x} \to \mathrm{b}_{K-1}...\mathrm{b}_1\mathrm{b}_0 \to z$.\footnote{
  Since the variables $x_i$ do not have an order, the map
  $\mathbf{x} \to \mathrm{b}_{K-1}...\mathrm{b}_1\mathrm{b}_0 \to z$ is not well
  defined unless we fix one.
  Given that any fixed order would be arbitrary, proximity of two integers, e.g.
  $z_1$ and $z_2 = z_1 + 1$, does not imply any relation between the values of
  $f(\mathbf{x}_1)$ and $f(\mathbf{x}_2)$.}

QSERA can find the extrema or roots of $f(\mathbf{x})$ by leveraging QUSA.
The first step of QSERA is to map extrema conditions,
$f(\mathbf{x}_*) = f_\mathrm{extr}$, or root conditions, $f(\mathbf{x}_*) = 0$,
to a function $u(z) \in \{0,1\}$, such that $u(z_*) = 1$ and
$u(z \neq z_*) = 0$, where $z_*$ corresponds to $\mathbf{x}_*$.
The second step is to find $z_*$, and thus $\mathbf{x}_*$, using QUSA.

The transformation $f(\mathbf{x}) \to u(z)$ requires the a priori knowledge of
the values of the global minimum ($f_{\min}$) and maximum ($f_{\max}$) because
we first need to rescale $f(\mathbf{x}) \to g(\mathbf{x})$, such that
$g(\mathbf{x}) \in [0,1]$.
This rescaling is slightly different depending on the problem; for maximisation
problems, we define:
\begin{align}
  g(\mathbf{x}) &= \frac{f(\mathbf{x}) - f_{\min}}{f_{\max} - f_{\min}}\,,
  \label{eq:g_max}
\end{align}
for minimisation problems:
\begin{align}
  g(\mathbf{x}) &= \frac{f_{\max} - f(\mathbf{x})}{f_{\max} - f_{\min}}\,,
  \label{eq:g_min}
\end{align}
and for root finding problems:
\begin{align}
  g(\mathbf{x}) &= 1 - \frac{f(\mathbf{x})^2}{\max(f_{\min}^2,f_{\max}^2)}\,.
  \label{eq:g_root}
\end{align}
Notice that in all three cases $g(\mathbf{x}_*) = 1$ and
$0 \leq g(\mathbf{x} \neq \mathbf{x}_*) < 1$.
We can then transform $g(\mathbf{x}) \to u(z)$ by raising $g(\mathbf{x})$ to the
power $n \in \mathbb{N}$:
\begin{align}
  u(z) &\equiv \lim_{n\to\infty}g(\mathbf{x})^n\,,
\end{align}
which essentially gives $u(z_*) = 1$ when $f(\mathbf{x}_*) = f_{\max}$,
$f(\mathbf{x}_*) = f_{\min}$, or $f(\mathbf{x}_*) = 0$, respectively for each
case, and $u(z) = 0$ for all other $z$.
Since in practical applications $n$ needs to be finite, we can define:
\begin{align}
  u_n(z) &\equiv g(\mathbf{x})^n\,,
\label{eq:u_n}
\end{align}
where the higher the value of $n$ the better $u_n(z)$ approximates $u(z)$.

The function $g(\mathbf{x})$ can be written as:
\begin{align}
g(\mathbf{x})
&= c_1 + \sum_{i}c_{1,i}x_i \nonumber\\
  &\quad+ \sum_{i_1}\sum_{i_2>i_1}c_{1,i_1i_2}x_{i_1}x_{i_2}
  + \cdots\,,
\label{eq:g}
\end{align}
where $c_1$, $c_{1,i}$, $c_{1,i_1i_2}$, ..., $c_{1,i_1...i_K}$ are constants
given by Eqs.~(\ref{eq:f_pairs}) and (\ref{eq:g_max})-(\ref{eq:g_root}).
Because $x_i^n = x_i$, note that $g(\mathbf{x})^n$ can be expressed as:
\begin{align}
g(\mathbf{x})^n
&= c_n + \sum_{i}c_{n,i}x_i \nonumber\\
  &\quad+ \sum_{i_1}\sum_{i_2>i_1}c_{n,i_1i_2}x_{i_1}x_{i_2}
  + \cdots\,,
\label{eq:gn}
\end{align}
where $c_n$, $c_{n,i}$, $c_{n,i_1i_2}$, ..., $c_{n,i_1...i_K}$ are constants
that depend on $c_1$, $c_{1,i}$, $c_{1,i_1i_2}$, ..., $c_{1n,i_1...i_K}$ of
Eq.~(\ref{eq:g}), and the subscript $n$ refers to the power of $g(\mathbf{x})$.

\begin{figure*}
\includegraphics[width=\textwidth]{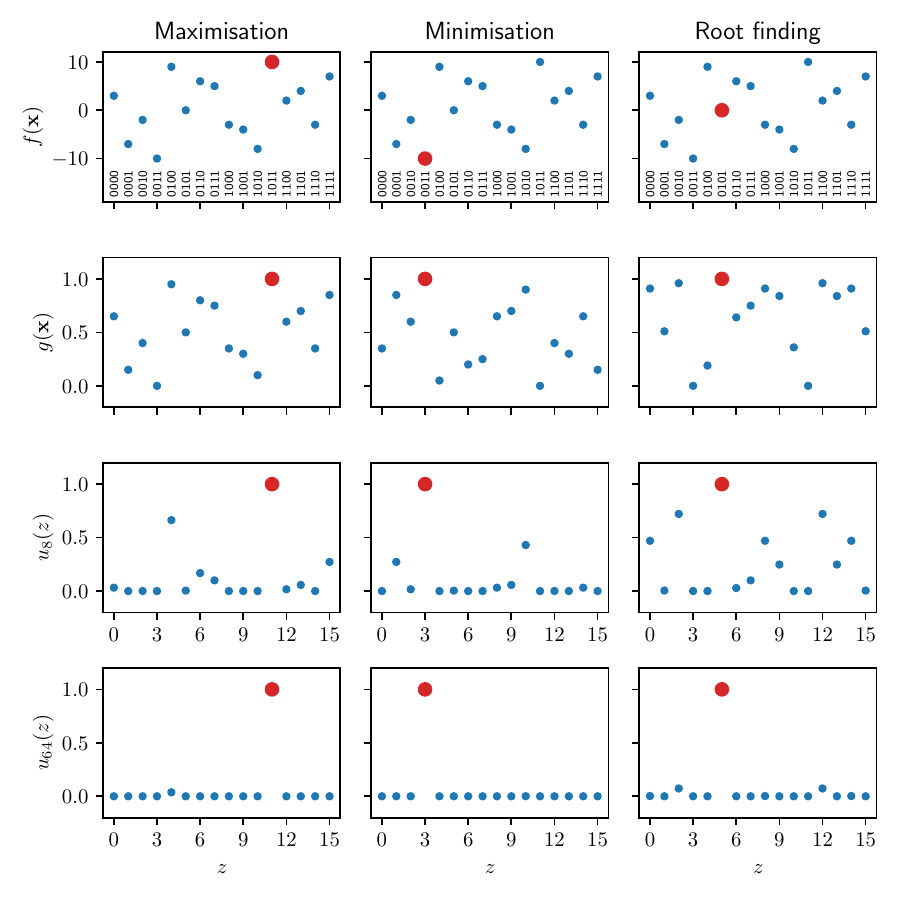}
\caption{
  Visualisation of the first step of QSERA, i.e. the transformation
  $f(\mathbf{x}) \to u(z)$ for maximisation (left column), minimisation (middle
  column), and root finding (right column) problems.
  The example discrete function $f(\mathbf{x})$ is the same in all three cases
  and depends on four binary variables, $x_i\in\{0,1\}$, that can be mapped to
  integers with $z \equiv \mathrm{b}_3\mathrm{b}_2\mathrm{b}_1\mathrm{b}_0$,
  where the right hand side has the elements of the vector $\mathbf{x}$ written
  as a binary number.
  The first row shows $f(\mathbf{x})$ with blue points, whereas the maximum
  (left), minimum (middle), and root (right) are shown with red points.
  The second row shows the transformation $f(\mathbf{x}) \to g(\mathbf{x})$
  based on Eq.~(\ref{eq:g_max}) (left), Eq.~(\ref{eq:g_min}) (middle), and
  Eq.~(\ref{eq:g_root}) (right), respectively.
  The third and fourth rows show the transformation $g(\mathbf{x}) \to u_n(z)$
  of Eq.~(\ref{eq:u_n}) for $n = 8$ and $n = 64$, respectively.
  \label{fig:transformation_examples}}
\end{figure*}
Figure~\ref{fig:transformation_examples} visualises how the first step of QSERA
works by showing an example discrete function $f(\mathbf{x})$ that depends on
$4$ binary variables (first row), the corresponding scaled function
$g(\mathbf{x})$ (second row), and the functions $u_8(z)$ (third row) and
$u_{64}(z)$ (fourth row) for the three cases: maximisation (left column),
minimisation (middle column), and root finding problems (right column).

\subsection{Implementation of the quantum oracle $u_n(z)$ with quantum gates}

For the representation of the oracle $u_n(z)$ as a quantum gate, we start by
defining the self-adjoint operator $\mathcal{B}_k$ which has the property:
\begin{align}
\mathcal{B}_k\ket{\mathrm{b}_k} = x_k\ket{\mathrm{b}_k}\,,
\label{eq:b}
\end{align}
i.e. $\mathcal{B}\ket{0} = 0\ket{0}$ and $\mathcal{B}\ket{1} = 1\ket{1}$.
Since $Z\ket{0} = \ket{0}$ and $Z\ket{1} = -\ket{1}$, $\mathcal{B}_k$ can be
defined as:
\begin{align}
\mathcal{B}_k \equiv \frac{1}{2}(I - Z_k)\,,
\end{align}
and see App.~\ref{app:commutation} for additional properties.

Then, we can define the operator $\mathcal{G}$ acting on a state
$\ket{z} = \ket{\mathrm{b}_{K-1}...\mathrm{b}_1\mathrm{b}_0}$ such that:
\begin{align}
  \mathcal{G}\ket{z} = g(\mathbf{x})\ket{z}\,,
\end{align}
where $g(\mathbf{x})$ is an eigenvalue of $\mathcal{G}$ and $\ket{z}$ an
eigenstate, with $\braket{z_1}{z_2} = \delta_{z_1z_2}$, i.e. the states form an
orthonormal basis.
By applying $\mathcal{G}$ multiple times, it follows that 
\begin{align}
  \mathcal{G}^n\ket{z} = g(\mathbf{x})^n\ket{z}\,.
\end{align}
The operator $\mathcal{G}^n$ can be defined based on Eqs.~(\ref{eq:gn}) and
(\ref{eq:b}):
\begin{align}
\mathcal{G}^n = \sum_{j=0}^{K}\mathcal{C}_{n,j}\,,
\end{align}
where
\begin{align}
\mathcal{C}_{n,0} &= c_nI\,, \nonumber\\
\mathcal{C}_{n,1} &= \sum_kc_{n,k}\mathcal{B}_k\,, \nonumber\\
\mathcal{C}_{n,2} &= \sum_{k_1}\sum_{k_2>k_1}c_{n,k_1k_2}
  \mathcal{B}_{k_1}\mathcal{B}_{k_2}\,, \nonumber\\
&\,\,\,\vdots \label{eq:Cj} \\
\mathcal{C}_{n,K}
  &= \sum_{k_1}\!\!\sum_{k_2>k_1}\cdots\!\!\!\!\!\!\!\!
    \sum_{k_K>k_1,...,k_{K-1}}
    \!\!\!\!\!\!\!\!\!\!\!c_{n,k_1k_2...k_K}\mathcal{B}_{k_1}\mathcal{B}_{k_2}\!
    \cdots \mathcal{B}_{k_K}\,. \nonumber
\end{align}

This enables us to define the unitary operator
\begin{align}
\mathcal{Q}_*
&\equiv e^{i\pi\mathcal{G}^n}
= e^{i\pi\sum_j\mathcal{C}_j}
= \prod_{j=0}^K\mathcal{Q}_{*j}\,,
\label{eq:qstar}
\end{align}
where $\mathcal{Q}_{*j} = e^{i\pi\mathcal{C}_j}$ and we dropped the subscript
$n$ from $\mathcal{Q}_* \equiv \mathcal{Q}_{*n}$,
$\mathcal{Q}_{*j} \equiv \mathcal{Q}_{*n,j}$,
$\mathcal{C}_j \equiv \mathcal{C}_{n,j}$, $c \equiv c_n$, $c_k \equiv c_{n,k}$,
$c_{k_1k_2} \equiv c_{n,k_1k_2}$, ..., $c_{k_1...k_K} \equiv c_{n,k_1...k_K}$
to simplify notation.
For $n \to \infty$, note that Eq.~(\ref{eq:qstar}) is a reflection operator
identical to that of Eq.~(\ref{eq:reflection}), see App.~\ref{app:reflection}.

For the zeroth order term, $j = 0$, we have $\mathcal{C}_0\ket{z} = c\ket{z}$
and
\begin{align}
\mathcal{Q}_{*0}\ket{z} &= e^{i\phi_0}\ket{z}\,,
\label{eq:Q_0}
\end{align}
where $\phi_0 = c\pi$, see App.~\ref{app:Q0_operator}.
Since $\phi_0$ does not depend on $z$, $\mathcal{Q}_{*0}$ can be implemented
with the help of an ancilla qubit and the phase shift gate, $P(\phi_0)$, by
rewriting the right-hand side of Eq.~(\ref{eq:Q_0}) as:
\begin{align}
\ket{z}\otimes e^{i\phi_0}\ket{1}_\mathrm{anc}
&= \ket{z}\otimes P(\phi_0)\ket{1}_\mathrm{anc}\,.
\end{align}
The quantum gate of the operator $\mathcal{Q}_{*0}$ is thus:
\begin{center}
\begin{quantikz}
\lstick{$\ket{z}\quad\,$}
  & \qwbundle[alternate]{}\gategroup[2, steps=3,
    style={dashed, rounded corners, inner xsep=0}]{$\mathcal{Q}_{*0}$}
  & \qwbundle[alternate]{}
  & \qwbundle[alternate]{}
  & \qwbundle[alternate]{}\rstick{$\ket{z}$} \\
\lstick{$\ket{0}_\mathrm{anc}$}
  & \gate{X}
  & \gate{P(\phi_0)}
  & \gate{X}
  & \qw\rstick{$\ket{0}_\mathrm{anc}$}
\end{quantikz}
\end{center}
For the first order terms, the operator $\mathcal{C}_1$ gives:
\begin{align}
\mathcal{Q}_{*1}\ket{z}
&= \prod_{k=0}^{K-1}\mathcal{P}_{1,k}(\phi_{1,k})\ket{z}\,,
\end{align}
where $\phi_{1,k} = c_k\pi$ and the quantum gate $\mathcal{P}_{1,k}(\phi_{1,k})$
is:
\begin{center}
\begin{quantikz}[row sep=0.4cm]
\lstick{$\ket{b_0}\quad\,$}
  & \qw\gategroup[6,steps=3,style={dashed, rounded corners}]
    {$\mathcal{P}_{1,k}(\phi_{1,k})$}
  & \qw
  & \qw
  & \qw\rstick{$\ket{b_0}$} \\
\wave&&&& \\
\lstick{$\ket{b_k}\quad\,$}
  & \ctrl{3}
  & \qw
  & \ctrl{3}
  & \qw\rstick{$\ket{b_k}$} \\
\wave&&&& \\
\lstick{$\ket{b_{K\!-\!1}}$}
  & \qw
  & \qw
  & \qw
  & \qw\rstick{$\ket{b_{K\!-\!1}}$} \\
\lstick{$\ket{0}_\mathrm{anc}\,$}
  & \targ{}
  & \gate{P(\phi_{1,k})}
  & \targ{}
  & \qw\rstick{$e^{i\phi_{1,k}}\ket{0}_\mathrm{anc}$}
\end{quantikz}
\end{center}
see App.~\ref{app:Q1_operator}.
For the second-order terms, the operator $\mathcal{Q}_{*2}$ is:
\begin{align}
\mathcal{Q}_{*2}\ket{z}
&= \prod_{k_1=0}^{K-1}\prod_{k_2>k_1}^{K-1}
  \mathcal{P}_{2,k_1k_2}(\phi_{2,k_1k_2})\ket{z}\,,
\end{align}
where $\phi_{2,k_1k_2} = c_{k_1k_2}\pi$ and the quantum gate
$\mathcal{P}_{2,k_1k_2}(\phi_{2,k_1k_2})$ is:
\begin{center}
\begin{quantikz}[row sep=0.4cm, column sep=0.4cm]
\lstick{$\ket{b_0}\quad\,$}
  & \qw\gategroup[8,steps=3,style={dashed, rounded corners}]
    {$\mathcal{P}_{2,k_1k_2}(\phi_{2,k_1k_2})$}
  & \qw
  & \qw
  & \qw\rstick{$\ket{b_0}$} \\
\wave&&&& \\
\lstick{$\ket{b_{k_1}}\,\,\,\,$}
  & \ctrl{2}
  & \qw
  & \ctrl{2}
  & \qw\rstick{$\ket{b_{k_1}}$} \\
\wave&&&& \\
\lstick{$\ket{b_{k_2}}\,\,\,\,$}
  & \ctrl{3}
  & \qw
  & \ctrl{3}
  & \qw\rstick{$\ket{b_{k_2}}$} \\
\wave&&&& \\
\lstick{$\ket{b_{K\!-\!1}}$}
  & \qw
  & \qw
  & \qw
  & \qw\rstick{$\ket{b_{K\!-\!1}}$} \\
\lstick{$\ket{0}_\mathrm{anc}\,$}
  & \targ{}
  & \gate{P(\phi_{2,k_1k_2})}
  & \targ{}
  & \qw\rstick{$e^{i\phi_{2,k_1k_2}}\ket{0}_\mathrm{anc}$}
\end{quantikz}
\end{center}
see App.~\ref{app:Q2_operator}.
In App.~\ref{app:Qn_operator} we show how to generalise this approach to derive
operators for $j$-th order terms:
\begin{align}
\mathcal{Q}_{*j}
  = \prod_{k_1}\,\cdots\!\!\!\!\!\!\!\!\!
  \prod_{k_j>k_1,...k_{j-1}}\!\!\!\!\!\!\!\!\!
  \mathcal{P}_j(\phi_{j,k_1...k_j})\ket{\mathrm{b}_{K-1}...\mathrm{b}_0}\,,
\end{align}
where $\mathcal{P}_j(\phi_{j,k_1...k_j})$ is a gate that adds the phase
$\phi_{j,k_1...k_,} = c_{k_1...k_j}\pi$ if $b_{k_1} = \cdots = b_{k_j} = 1$.
Bringing all these operators together, the gate
$\mathcal{Q}_* = \prod_j\mathcal{Q}_{*j}$ is:
\begin{center}
$\begin{quantikz}[column sep=0.2cm]
\qwbundle[alternate]{}
  & \gate{\mathcal{Q}_*}\qwbundle[alternate]{}
  & \qwbundle[alternate]{}
\end{quantikz}\equiv\begin{quantikz}[column sep=0.2cm]
\qwbundle[alternate]{}
  & \gate{\mathcal{Q}_{*0}}\gategroup[1, steps=4,
    style={dashed, rounded corners, inner xsep=0}]{$\mathcal{Q}_*$}
    \qwbundle[alternate]{}
  & \gate{\mathcal{Q}_{*1}}\qwbundle[alternate]{}
  & \cdots\qwbundle[alternate]{}
  & \gate{\mathcal{Q}_{*j}}\qwbundle[alternate]{}
  & \qwbundle[alternate]{}
\end{quantikz}$
\end{center}

\section{Remarks}
\label{sec:discussion}

If the values $f_\mathrm{min}$ and $f_\mathrm{max}$ are known and the power $n$
is sufficiently large, QSERA is guaranteed to identify $\mathbf{x}_*$ after
$\mathcal{O}(\sqrt{N})$ iterations in a single quantum execution (single shot).
If these values are not known, however, appropriate estimates would be needed
and the probability of finding $\mathbf{x}_*$ would be less than $1$.
For example, consider a minimisation problem where the theoretical minimum of
the objective function is zero, but none of the $\mathbf{x}$ exactly gives that
value, i.e. $f(\mathbf{x}_*) = f_\mathrm{min} > 0$ (a specific example of this
is studied in Sect.~\ref{sec:application}).
Setting estimated values for $f_\mathrm{min}$ and $f_\mathrm{max}$ in
Eq.~(\ref{eq:g_min}) that are less/greater than the true minimum/maximum of
$f(\mathbf{x})$, respectively, may still allow QSERA to work.
However, in this case $g(\mathbf{x}_*) < 1$ and thus
$g(\mathbf{x}_*)^n = u_n(z_*) < 1$, leading to
$u(z) = \lim_{n\to\infty}u_n(z) \to 0$ for all $z$.
This implies that the oracle loses its ability to identify $z_*$, which is the
opposite behaviour from the case in which the values $f_\mathrm{min}$,
$f_\mathrm{max}$ are known and the probability of measuring $\mathbf{x}_*$ goes
to $1$ as $n$ increases.
Nevertheless, there should be an optimal value of $n = n_\mathrm{opt}$ that is
large enough to sufficiently identify the points near the optimum, but not too
large to send $u_n(z_*)$ to $0$ (see Sect.~\ref{sec:application}).

Another potential limitation of QSERA is that the parameters $c_n$, $c_{n,i}$,
$c_{n,i_1i_2}$, ... --- needed to define the quantum gates $\mathcal{C}_j$
(Eq.~\ref{eq:Cj}) --- may not be easy to calculate.
To derive them, start from the known parameters $w$, $w_i$, $w_{i_1i_2}$, ... of
$f(\mathbf{x})$ and the corresponding parameters of $g(\mathbf{x})$ --- $c_1$,
$c_{1,i}$, $c_{1,i_1i_2}$, ... --- which, for optimisation problems, are just a
rescaled version of the former.
Then, the parameters $c_n$, $c_{n,i}$, $c_{n,i_1i_2}$, ... of $g(\mathbf{x})^n$
can be recursively calculated with a classical computer based on
$g(\mathbf{x})^n = g(\mathbf{x})^{n-1}g(\mathbf{x})$.
However, while this should be in principle feasible, it may not be trivial to do
for large $N$.

Apart from identifying roots and extrema of discrete functions, QSERA is general
enough that can also work for continuous functions, e.g. $h(y)$ with
$y\in\mathbb{R}$.
For such functions, an initial step is required to discretise $h(y)$ with $N$
points, i.e. to create pairs of $y_i$ and $h(y_i)$ in order to map $y_i$ to the
integers $z$ and $h(y_i)$ to $u(z)$.
The larger the number of points the closer the result will be to the optimum
$y_*$, but at the same time the deeper the quantum circuit and the larger the
number of QUSA iterations needed.
Since structured search (e.g. gradient descent) is generally expected to be more
efficient for functions with continuous variables, QSERA could be useful in
cases where these approaches may fail, such as when $h(y)$ is ``noisy'' and/or
has steep gradients, properties which can trap gradient descent algorithms into
local extrema.

\section{Application of QSERA to portfolio optimisation}
\label{sec:application}

The practical implementation of QSERA can be described with the following
portfolio optimisation problem.
Consider $N_\mathrm{a} = 4$ assets, labeled \textsc{a}, \textsc{b}, \textsc{c},
and \textsc{d}, with the following mean returns, volatility, and correlations:
\begin{align}
\mu &= \left[\begin{array}{cccc}
0.05 & 0.01 & 0.02 & 0.04
\end{array}\right]\,,
\end{align}
\begin{align}
\sigma &= \left[\begin{array}{cccc}
0.40 & 0.10 & 0.20 & 0.30
\end{array}\right]\,,
\end{align}
\begin{align}
\rho &= \left[\begin{array}{r@{.}lr@{.}lr@{.}lr@{.}l}
 1&0 &  0&5 & -0&4 & -0&2 \\
 0&5 &  1&0 & -0&1 & -0&3 \\
-0&4 & -0&1 &  1&0 &  0&3 \\
-0&2 & -0&3 &  0&3 &  1&0
\end{array}\right]\,.
\end{align}
In the following, we label portfolios with the binary number:
$\mathrm{b}_\textsc{d}\mathrm{b}_\textsc{c}
\mathrm{b}_\textsc{b}\mathrm{b}_\textsc{a}$.\footnote{
  For example, the portfolio \texttt{0101} has
  $x_\textsc{a} = x_\textsc{c} = 1$, $x_\textsc{b} = x_\textsc{d} = 0$,
  $\mu_\mathrm{p} = (\mu_\textsc{a} + \mu_\textsc{c})/2$ and
  $\sigma_\mathrm{p}^2 = (\sigma_\textsc{a}^2 + \sigma_\textsc{c}^2
    + 2\sigma_\textsc{a}\sigma_\textsc{c}\rho_\textsc{ac})/4$.}

The objective function depends on the investor goals and can take many different
forms.
For this example, assume that we want to construct a portfolio that tracks a
benchmark with $\mu_\mathrm{b} = 0.043$ and $\sigma_\mathrm{b} = 0.195$, with a
budget that corresponds to $N_\mathrm{b} = 2$ assets.
We can represent this as a minimisation problem,
$f(x_{\textsc{a}*}, x_{\textsc{b}*}, x_{\textsc{c}*}, x_{\textsc{d}*})
= f_\mathrm{min}$, with the objective function:
\begin{align}
&f(x_\textsc{a}, x_\textsc{b}, x_\textsc{c}, x_\textsc{d}) = \nonumber\\
&= \lambda_\mu
  \left(N_\mathrm{p}\mu_\mathrm{p} - N_\mathrm{p}\mu_\mathrm{b}\right)^2
  + \lambda_{\sigma^2}\left(N_\mathrm{p}^2\sigma_\mathrm{p}^2
    - N_\mathrm{p}^2\sigma_\mathrm{b}^2\right)^2
  \nonumber\\
&\quad+ (1 - \lambda_\mu - \lambda_{\sigma^2})
  \left(N_\mathrm{p} - N_\mathrm{b}\right)^2\,,
\label{eq:f}
\end{align}
where $N_\mathrm{p} = x_\textsc{a} + x_\textsc{b} + x_\textsc{c}+ x_\textsc{d}$,
$\mu_\mathrm{p}$ and $\sigma_\mathrm{p}$ are given by Eqs.~(\ref{eq:mu_p}) and
(\ref{eq:sigma_p}), respectively, and the parameters
$\lambda_\mu, \lambda_{\sigma^2} \in [0,1]$ (with
$\lambda_\mu + \lambda_{\sigma^2} \leq 1$) control whether the portfolio is
constructed to more closely track the returns or volatility of the benhchmark;
we set $\lambda_\mu = 0.95$ and $\lambda_{\sigma^2} = 0.049$.
Here, the budget has been introduced as a ``soft'' constraint (last term of
Eq.~\ref{eq:f}) with the multiplier:
$1 - \lambda_\mu - \lambda_{\sigma^2} = 0.001$.

\begin{figure}
\includegraphics[width=\columnwidth]{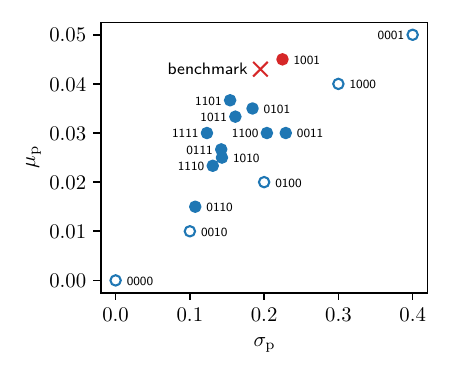}
\caption{
  The returns and volatility of the $2^{N_\mathrm{a}} = 16$ portfolios that can
  be constructed from the combinations of the $N_\mathrm{a} = 4$ available
  assets.
  The \texttt{0000} portfolio and the portfolios consisting of one asset only
  are shown with open blue points, those with two or more assets with filled
  blue points, the benchmark with a red cross, and the portfolio that minimises
  $f(\mathbf{x})$ --- which QSERA is expected to find --- with a red point.}
\label{fig:portfolio_combinations}
\end{figure}
Figure~\ref{fig:portfolio_combinations} shows $\mu_\mathrm{p}$ and
$\sigma_\mathrm{p}$ for the $2^{N_\mathrm{a}} = 16$ portfolios (blue points)
alongside the benchmark (red cross), and notice that no portfolio exactly
matches the values $\mu_\mathrm{b}$ and $\sigma_\mathrm{b}$.
QSERA is expected to find the portfolio \texttt{1001} (shown with a red point)
which minimises $f(\mathbf{x})$ because it is the closest to the benchmark and
has $N_\mathrm{p} = N_\mathrm{b}$.
Note that the portfolio \texttt{0101} is also relatively close to the red cross,
albeit farther, and consists of two assets as well; thus, its corresponding
value of the objective function is expected to be near the minimum.

\begin{figure}
\includegraphics[width=\columnwidth]{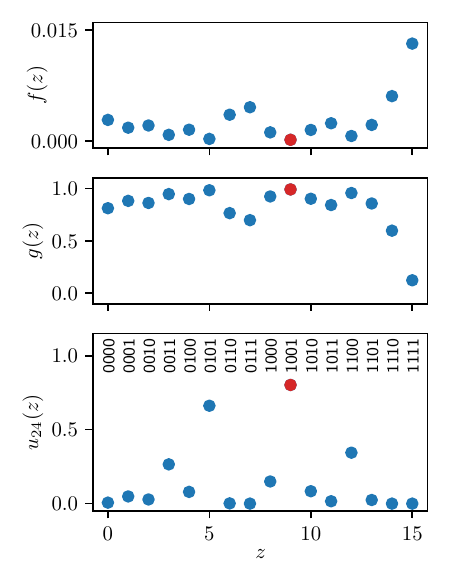}
\caption{
  The functions $f(z)$ (Eq.~\ref{eq:f}, top), $g(z)$ (Eq.~\ref{eq:g_min} with
  $f_\mathrm{min}$ set to $0$, middle), and $u_{24}(z)$ (Eq.~\ref{eq:u_n},
  bottom), which show how the quantum oracle is created.}
\label{fig:functions}
\end{figure}
Figure~\ref{fig:functions} shows the first step of QSERA,
$f(\mathbf{x}) \to u(z)$; namely, how the objective function (top panel) is
transformed to an oracle (bottom panel) that can identify $z_*$ (and hence
$\mathbf{x}_*$).
The value of $f(\mathbf{x_*})$ is not known a priori because none of the
available portfolios make Eq.~(\ref{eq:f}) equal to zero, which is the
theoretical minimum.
However, the actual minimum is expected to be close to zero, thus we set
$f_\mathrm{min} \simeq 0$ in the definition of $g(\mathbf{x})$ of
Eq.~(\ref{eq:g_min}).
Similarly, we set $f_\mathrm{max} \simeq 0.15$ in the same definition, which is
larger than the true maximum of $f(\mathbf{x})$ but which is not known either.
These lead to $g(\mathbf{x_*}) < 1$, and therefore
$g(\mathbf{x}_*)^n = u_n(z_*) < 1$; this is the case discussed in
Sect.~\ref{sec:discussion}.

\begin{table}
\begin{center}
\begin{tabular}{cr@{.}l|cr@{.}l|cr@{.}l}
\hline
\multicolumn{3}{r|}{$f(\mathbf{x})\ \times\!\!10^{-3}\!\!$}
  & $g(\mathbf{x})$ & $\times$0&1
  & $u_{24}(z)$ & \multicolumn{2}{c}{$\times$1} \\
\hline
\hline
$w$               &  2&8 & $c_1$                  &  8&1
                         & $c_{24}$               &  0&0 \\
\hline
$w_\textsc{a}$    & -1&1 & $c_{1,\textsc{a}}$     &  0&7
                         & $c_{24,\textsc{a}}$    &  0&0 \\
$w_\textsc{b}$    & -0&8 & $c_{1,\textsc{b}}$     &  0&5
                         & $c_{24,\textsc{b}}$    &  0&0 \\
$w_\textsc{c}$    & -1&3 & $c_{1,\textsc{c}}$     &  0&9
                         & $c_{24,\textsc{c}}$    &  0&1 \\
$w_\textsc{d}$    & -1&7 & $c_{1,\textsc{d}}$     &  1&1
                         & $c_{24,\textsc{d}}$    &  0&1 \\
\hline
$w_\textsc{ab}$   & -0&2 & $c_{1,\textsc{ab}}$    &  0&1
                         & $c_{24,\textsc{ab}}$   &  0&2 \\
$w_\textsc{ac}$   & -0&2 & $c_{1,\textsc{ac}}$    &  0&1
                         & $c_{24,\textsc{ac}}$   &  0&5 \\
$w_\textsc{ad}$   &  0&0 & $c_{1,\textsc{ad}}$    &  0&0
                         & $c_{24,\textsc{ad}}$   &  0&6 \\
$w_\textsc{bc}$   &  2&8 & $c_{1,\textsc{bc}}$    & -1&9
                         & $c_{24,\textsc{bc}}$   & -0&1 \\
$w_\textsc{bd}$   &  1&1 & $c_{1,\textsc{bd}}$    & -0&7
                         & $c_{24,\textsc{bd}}$   & -0&1 \\
$w_\textsc{cd}$   &  0&8 & $c_{1,\textsc{cd}}$    & -0&6
                         & $c_{24,\textsc{cd}}$   &  0&1 \\
\hline
$w_\textsc{abc}$  &  2&5 & $c_{1,\textsc{abc}}$   & -1&6
                         & $c_{24,\textsc{abc}}$  & -0&8 \\
$w_\textsc{abd}$  &  2&1 & $c_{1,\textsc{abd}}$   & -1&4
                         & $c_{24,\textsc{abd}}$  & -0&9 \\
$w_\textsc{acd}$  &  2&7 & $c_{1,\textsc{acd}}$   & -1&8
                         & $c_{24,\textsc{acd}}$  & -1&5 \\
$w_\textsc{bcd}$  &  2&3 & $c_{1,\textsc{bcd}}$   & -1&5
                         & $c_{24,\textsc{bcd}}$  & -0&2 \\
\hline
$w_\textsc{abcd}$ &  3&9 & $c_{1,\textsc{abcd}}$  & -0&8
                         & $c_{24,\textsc{abcd}}$ &  1&8 \\
\hline
\end{tabular}
\end{center}
\caption{
  The (rounded) parameters of the zeroth, first, second, third, and fourth order
  terms of $f(\mathbf{x})$, $g(\mathbf{x})$, and $u_{24}(z)$.}
\label{tab:parameters}
\end{table}
The objective function $f(\mathbf{x})$ (Eq.~\ref{eq:f}) can be written in the
form of Eq.~(\ref{eq:f_pairs}) and, in turn, $g(\mathbf{x})$
(Eq.~\ref{eq:g_min}) can be written in the form of Eq.~(\ref{eq:g}), see
App.~\ref{app:f}.
Then, the parameters of Eq.~(\ref{eq:gn}), which are needed to define the
operators $\mathcal{C}_j$ (Eq.~\ref{eq:Cj}), can be recursively calculated based
on $g(\mathbf{x})^n = g(\mathbf{x})^{n-1}g(\mathbf{x})$.
Following Eq.~(\ref{eq:gn}), this process gives the parameters $c_n$, $c_{n,i}$,
$c_{n,i_1i_2}$, ..., $c_{n,i_1...i_K}$ as a function of $c_1$, $c_{1,i}$,
$c_{1,i_1i_2}$, ..., $c_{1n,i_1...i_K}$ of Eq.~(\ref{eq:g}).
Table~\ref{tab:parameters} lists these parameters for the zeroth to the fourth
order terms, which are then used to assemble the quantum gates
$\mathcal{Q}_{*j}$.

For the quantum circuit, we use four qubits to represent the $N = 2^4$
portfolios as quantum states, $\ket{\mathrm{b}_\textsc{d}\mathrm{b}_\textsc{c}
\mathrm{b}_\textsc{b}\mathrm{b}_\textsc{a}}$, two ancilla qubits to implemement
the $\textsc{and}$ operators, $\ket{00}_\textsc{and}$, and one ancilla qubit to
track the phase of the states, $\ket{0}_\mathrm{anc}$.
Appendix~\ref{app:circuit} shows the decomposition of the quantum gates
$\mathcal{Q}_H$, $\mathcal{Q}_*$, and $\mathcal{Q}_+$, down to the basic quantum
gates: $H$, $X$, $Z$, $P$, and Toffoli.
Since the number of portfolio combinations is $N = 16$, the optimal value of the
QUSA iterations is $m_\mathrm{opt} = 2$.

\begin{figure}
\includegraphics[width=\columnwidth]{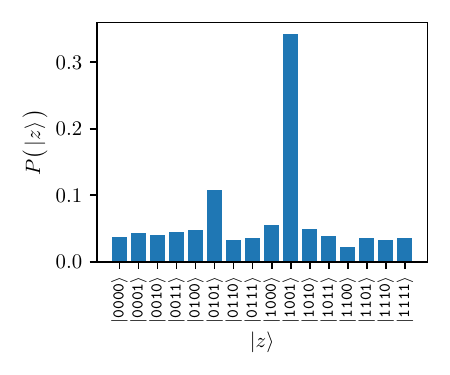}
\caption{
  The probabilities of measuring the different states, each one representing a
  portfolio.
  QSERA identifies the portfolio \texttt{1001} which minimises the objective
  function of Eq.~(\ref{eq:f}).
  The portfolio \texttt{0101}, which has the second lowest value, also has a
  high probability to be measured relative to all other ones.}
\label{fig:measurement}
\end{figure}
Figure~\ref{fig:measurement} shows the probability distribution of measuring
the different states after applying the quantum gates
$(\mathcal{Q}_+\mathcal{Q}_*)^2\mathcal{Q}_H$ to the initial state (the pair
$\mathcal{Q}_+\mathcal{Q}_*$ is applied twice because $m_\mathrm{opt} = 2$).
The portfolio \texttt{1001}, which is the solution of the optimisation problem,
has the highest probability of being measured.
As discussed in Sect.~\ref{sec:discussion}, its probability is less than $1$
because we used approximations for $f_\mathrm{min}$ and $f_\mathrm{max}$ in the
definition of $g(\mathbf{x})$.
Notice that the portfolio \texttt{0101} is the second most likely to be measured
because it is the second closest to the benchmark with
$N_\mathrm{p} = N_\mathrm{b} = 2$.
This suggests that when $f_\mathrm{min}$, $f_\mathrm{max}$ are approximated, the
result of QSERA of a single shot may not be the actual optimum but a point close
to it.
Multiple shots would be needed to identify the true optimum.

\begin{figure}
\includegraphics[width=\columnwidth]{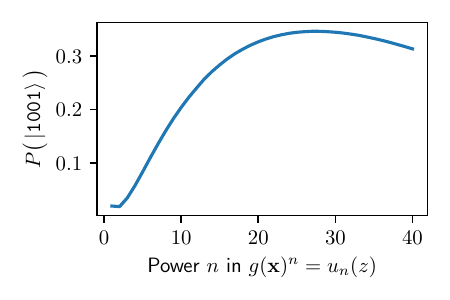}
\caption{
  The probability of measuring the portfolio that minimises $f(\mathbf{x})$ as
  a function of the power $n$ used to create the oracle $u_n(z)$.}
\label{fig:n_dependence}
\end{figure}

Figure~\ref{fig:n_dependence} shows how the probability of measuring
$\ket{\texttt{1001}}$ changes as the power $n$ increases.
For small values of $n$, $u_n(z)$ is still quite similar to $g(z)$ (see middle
panel of Fig.~\ref{fig:functions}) and the oracle cannot distinguish the
portfolio that minimises $f(\mathbf{x})$ from other portfolios near the minimum.
For $n$ in the range $\sim$$20$-$30$, $u_n(z)$ is similar to $u_{24}(z)$ shown
on the bottom panel of Fig.~\ref{fig:functions}, and hence the quantum oracle
$\mathcal{Q}_*$ can effectively amplify the probability of measuring $z_*$.
For larger $n$, $u_n(z) \to 0$ for all $z$, and thus the probability of
measuring the optimal portfolio also goes to zero.
Essentially, Fig.~\ref{fig:n_dependence} is a visualisation of the corresponding
discussion in Sect.~\ref{sec:discussion}.

\section{Summary and conclusions}
\label{sec:conclusions}

We propose a quantum unstructured search algorithm, QSERA, that can find the
roots or extrema $\mathbf{x}_*$ of discrete functions $f(\mathbf{x})$, and can
thus be used to solve discrete optimisation problems such as combinatorial
optimisation.
QSERA consists of two steps: first, the objective function $f(\mathbf{x})$ is
mapped to a function $u(z)$ serving as an oracle that can identify the root or
extremum, and, second, Grover's algorithm is employed to find $z_*$ in
$\mathcal{O}(\sqrt{N})$ iterations.
In the paper, we provide the mathematical description of the algorithm,
demonstrate how to build the relevant quantum gates for its implementation, and
assemble a quantum circuit for portfolio optimisation to show how QSERA works in
practice.

QSERA can be applied to problems with objective functions of any order, and thus
is more general than the algorithms requiring the QUBO formulation.
QSERA can also be applied to continuous functions, which may prove useful when
structured search (e.g. gradient descent) is ineffective, such as when the
function is ``noisy'' and/or has steep gradients.
One of the limitations of the algorithm is that it can find the optimum
$\mathbf{x}_*$ with certainty only if $f_\mathrm{min}$ and $f_\mathrm{max}$ are
known a priori.
Otherwise, appropriate estimates should be provided, and while the state
representing $\mathbf{x}_*$ will be the most likely to be measured, multiple
shots would be needed to distinguish it from other points near the optimum.

\acknowledgements{
  The quantum computations were performed on a classical computer using a
  zero-noise simulator from Qiskit (version \texttt{1.0}), an open-source
  framework for quantum computing \cite{Qiskit}.
}

\bibliographystyle{quantum}
\bibliography{paper}

\appendix

\onecolumn

\section{Quantum optimisation algorithms}

\subsection{Quantum Adiabatic Algorithm (QAA)}
\label{app:qaa}

The QAA family of algorithms are based on the Schr\"odinger equation:
$i\frac{d}{dt}\ket{\psi(t)} = \mathcal{H}\ket{\psi(t)}$, where the operator
$\mathcal{H}$ is the Hamiltonian.
The eigenstates, $\ket{\psi_n}$, and eigenvalues, $E_n$, of $\mathcal{H}$
satisfy the equation: $\mathcal{H}\ket{\psi_n} = E_n\ket{\psi_n}$, with
$E_n^{-1}$ characteristic timescales.
We can define a time-evolution operator, $\mathcal{U}(t_0 + t, t_0)$, that takes
a state from time $t_0$ to a time $t_0 + t$:
$\ket{\psi(t_0 + t)} = \mathcal{U}(t_0 + t, t_0)\ket{\psi(t_0)}$, where
$\mathcal{U}(t_2, t_0) = \mathcal{U}(t_2, t_1)\mathcal{U}(t_1, t_0)$ for
$t_2 \geq t_1 \geq t_0$.
We can then rewrite the Schr\"odinger equation as:
$i\frac{d}{dt}\mathcal{U}(t_0 + t, t_0) = \mathcal{H}\mathcal{U}(t_0 + t, t_0)$,
which has the solution: $\mathcal{U}(t_0 + t, t_0) = e^{-i\mathcal{H}t}$.
If the Hamiltonian is time-dependent, but it varies much more slowly than the
timescales $E_n^{-1}$, then for $t$ on the order of $E_n^{-1}$ we can consider
$\mathcal{H}(t_0 + t)$ to be constant and the solution of the Schr\"odinger
equation can be approximated as:
$\mathcal{U}(t_0 + t, t_0) \simeq e^{-i\mathcal{H}(t_0 + t)t}$.

We can leverage this property to find the unknown ground state of a Hamiltonian
$\mathcal{H}_\mathrm{c}$ as follows: i) prepare the system in the ground state
of a known Hamiltonian, $\mathcal{H}_0$, ii) slowly evolve the Hamiltonian
towards $\mathcal{H}_\mathrm{c}$, and iii) measure the final state of the system
which will be the ground state of $\mathcal{H}_\mathrm{c}$.
In particular, consider the Hamiltonian:
$\mathcal{H}(t) = \frac{T-t}{T}\mathcal{H}_0+\frac{t}{T}\mathcal{H}_\mathrm{c}$,
where $\mathcal{H}_0 = \mathcal{H}(0)$ and
$\mathcal{H}_\mathrm{c} = \mathcal{H}(T)$ are time-independent, and the
eigenstates of $\mathcal{H}_0$ are known but those of $\mathcal{H}_\mathrm{c}$
are not.
The operator $\mathcal{H}_\mathrm{c}$ corresponds to the function we want to
minimise and its ground state to the global minimum.
For the practical implementation, we discretise the time interval $[0, T]$ with
$M$ timesteps, $\Delta t$, such that:
\begin{align}
\mathcal{U}(T, 0) &= \prod_{m=0}^{M-1}U(m\Delta t + \Delta t, m\Delta t)
\simeq \prod_me^{-i\mathcal{H}(m\Delta t + \Delta t)\Delta t}\,,
\label{eq:U}
\end{align}
where $\mathcal{H}(m\Delta t)
= \frac{M - m}{M}\mathcal{H}_0 + \frac{m}{M}\mathcal{H}_\mathrm{c}$.
However, because $\mathcal{H}_0$ and $\mathcal{H}_\mathrm{c}$ do not commute,
we need to use the Trotter product formula for the right-hand side of
Eq.~(\ref{eq:U}).

\subsection{Variational Quantum Eigensolvers (VQE)}
\label{app:vqe}

Consider a function $f(z)\in[0,1]$ and a quantum register of
$K = K_\mathrm{in} + K_\mathrm{out}$ qubits, where $K_\mathrm{in}$ qubits are
used to represent $z$ and $K_\mathrm{out}$ qubits to compute $f(z)$.
We can define a self-adjoint operator $\mathcal{F}$ acting on a state $\ket{z}$
such that $\mathcal{F}\ket{z} = f(z)\ket{z}$, where $f(z)$ is an eigenvalue of
$\mathcal{F}$ and $\ket{z}$ an eigenstate, with the states
$\braket{z_1}{z_2} = \delta_{z_1z_2}$ forming an orthonormal basis
($\delta_{z_1z_2}$ is the Kronecker delta).
The eigenvalue $f(z)$ can be measured with a Quantum Phase Estimation (QPE)
algorithm.
First, define the operator
$\mathcal{Q} \equiv e^{i\pi\mathcal{F}}$ which when applied $m$ times to an
eigenstate gives $\mathcal{Q}^m\ket{z} = e^{im\phi}\ket{z}$, where
$\phi = \pi f(z)$.
Then, prepare the input qubits in the state
$\mathcal{Z}\ket{0}^{\otimes K_\mathrm{in}} = \ket{z}$ with an appropriate
operator $\mathcal{Z}$, and the output qubits in the state
$\mathrm{QFT}\ket{0}^{\otimes K_\mathrm{out}}
= \ket{+}^{\otimes K_\mathrm{out}}$ using a Quantum Fourier Transform (QFT)
gate.
By applying $\mathcal{Q}^m$ as a control gate --- with the control qubits being
the input ones and the target qubits the output ones --- we can exploit phase
kickback to encode multiples of the angle $\phi$ on the phases of the output
qubits.
Finally, based on quantum interference, we can apply an inverse QFT
($\mathrm{QFT}^\dagger$) gate to directly read off the value of $f(z)$ from the
output qubits.
The corresponding quantum circuit is:
\begin{center}
\begin{quantikz}
\lstick{$\ket{0}^{\otimes K_\mathrm{in}}\,\,\,$}
  & \gate{\mathcal{Z}}\qwbundle[alternate]{}
  & \push{\,\,\ket{z}\,\,}\qwbundle[alternate]{}
  & \gate{\prod_{k=0}^{K_\mathrm{out}\!-\!1}\mathcal{Q}^{2^k}}
    \qwbundle[alternate]{}
  & \qwbundle[alternate]{}
  & \qwbundle[alternate]{}\rstick{$\ket{z}$} \\
\lstick{$\ket{0}^{\otimes K_\mathrm{out}}$}
  & \qwbundle[alternate]{}
  & \gate{\mathrm{QFT}}\qwbundle[alternate]{}
  & \ctrlbundle{-1}\qwbundle[alternate]{}
  & \gate{\mathrm{QFT}^\dagger}\qwbundle[alternate]{}
  & \qwbundle[alternate]{}\rstick{$\ket{(2^{K_\mathrm{out}}-1)f(z)}$}
\end{quantikz}
\end{center}
where
\begin{center}
\begin{quantikz}[column sep=0.3cm]
\lstick{$\ket{z}\quad\quad\,$}
  & \gate{\mathcal{Q}^{2^0}}\qwbundle[alternate]{}
    \gategroup[5, steps=4, style={dashed, rounded corners}]
    {$\prod_{k=0}^{K_\mathrm{out}-1}\mathcal{Q}^{2^k}$}
  & \gate{\mathcal{Q}^{2^1}}\qwbundle[alternate]{}
  & \cdots\qwbundle[alternate]{}
  & \gate{\mathcal{Q}^{2^{K_\mathrm{out}-1}}}\qwbundle[alternate]{}
  & \qwbundle[alternate]{} \\
\lstick{$\ket{0}_\mathrm{out}^0\quad\,$}
  & \ctrl{-1}
  & \qw
  & \cdots\qw
  & \qw
  & \qw \\
\lstick{$\ket{0}_\mathrm{out}^1\quad\,$}
  & \qw
  & \ctrl{-2}
  & \cdots\qw
  & \qw
  & \qw \\
\wave&&&&& \\
\lstick{$\ket{0}_\mathrm{out}^{K_\mathrm{out}\!-\!1}$}
  & \qw
  & \qw
  & \cdots\qw
  & \ctrl{-4}
  & \qw
\end{quantikz}$\quad\ $and$\quad$\begin{quantikz}[column sep=0.3cm]
\qwbundle[alternate]{}
  & \gate{\mathcal{Q}^m}\qwbundle[alternate]{}
  & \qwbundle[alternate]{} \\
\qw
  & \ctrl{-1}
  & \qw
\end{quantikz}$\ \equiv$\begin{quantikz}[column sep=0.3cm]
\qwbundle[alternate]{}
  & \gate{\mathcal{Q}}\qwbundle[alternate]{}
  & \gate{\mathcal{Q}}\qwbundle[alternate]{}
  & \cdots\qwbundle[alternate]{}
  & \gate{\mathcal{Q}}\qwbundle[alternate]{}
  & \qwbundle[alternate]{} \\
\qw
  & \ctrl{-1}
  & \ctrl{-1}
  & \cdots\qw
  & \ctrl{-1}
  & \qw
\end{quantikz}
\end{center}
e.g. see Ref.~\cite{MatsakosNield2024} for a detailed implementation of QPE.
Given a superposition $\ket{\psi} = \sum_{z=0}^{2^{K_\mathrm{in}}-1}a_z\ket{z}$,
where $p_z \equiv |a_z|^2 = \braket{z}{\psi}$ is the probability of measuring the
state $\ket{z}$, we can use QPE to measure the expected value
$\left<f\right> = \bra{\psi}\mathcal{F}\ket{\psi} = \sum_zp_zf(z)$ based on the
fact that $\mathcal{F}\ket{\psi} = \sum_za_zf(z)\ket{z}$.

With a quantum circuit available to compute $\left<f\right>$, the extrema can be
found with a variational approach.
In particular, VQE algorithms employ an iterative optimisation process as
follows: i) initialise the input state
$\mathcal{A}(\mathbf{h})\ket{0}^{\otimes K_\mathrm{in}}
= \ket{\psi(\mathbf{h})}$ with a parametrised quantum gate
$\mathcal{A}(\mathbf{h})$ where $\mathbf{h} = (h_1, h_2, ..., h_{M})$ is a set
of $M$ parameters, ii) estimate $\left<f\right>$ with a QPE circuit, iii) feed
the result to a classical computer to obtain updated parameters $\mathbf{h}$
such that the next input state gets closer to the optimum of $f$, and iv) repeat
until convergence.

\subsection{Quantum Approximate Optimisation Algorithm (QAOA)}
\label{app:qaoa}

QAOA is inspired by QAA, with the main difference being the parametrisation of
$\mathcal{U}$ with parameters $\mathbf{h}_0$ and $\mathbf{h}_\mathrm{c}$, i.e.
$\mathcal{U}(T,0) \to \mathcal{U}(\mathbf{h}_0,\mathbf{h}_\mathrm{c})$, such
that a variational approach can be used to achieve convergence.
Specifically, the system is initialised in the ground state, $\ket{\psi_0}$, of
a known Hamiltonian, $\mathcal{H}_0$, and evolves towards the ground state of
the target Hamiltonian, $\mathcal{H}_\mathrm{c}$, based on the evolution
operator:
\begin{align}
\ket{\psi(\mathbf{h}_0,\mathbf{h}_\mathrm{c})}
&= \mathcal{U}(\mathbf{h}_0,\mathbf{h}_\mathrm{c})\ket{\psi_0}
  \label{eq:qaoa}
= \prod_{m=1}^{M}\left(
    e^{-ih_{0m}\mathcal{H}_0}
    e^{-ih_{\mathrm{c}m}\mathcal{H}_\mathrm{c}}
  \right)\ket{\psi_0}\,,
\end{align}
where $\mathbf{h}_0 = (h_{01},h_{02},...,h_{0M})$ and $\mathbf{h}_\mathrm{c}
= (h_{\mathrm{c}1},h_{\mathrm{c}2},...,h_{\mathrm{c}M})$ are $2M$ parameters
that can be varied to obtain the ground state of $\mathcal{H}_\mathrm{c}$.

\section{QUSA worked example}
\label{app:qusa_example}

\begin{figure}
\centering
\includegraphics[width=\textwidth]{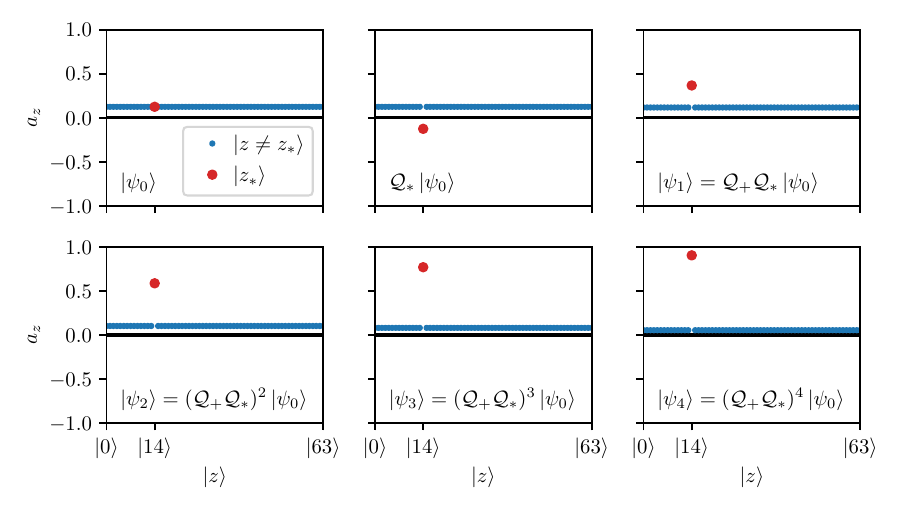}\\
\includegraphics[width=\textwidth]{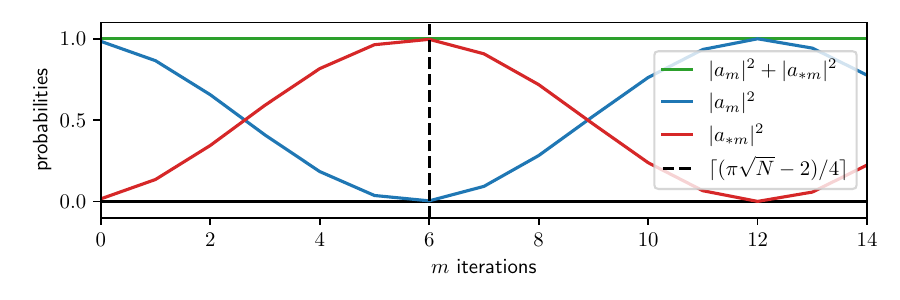}
\caption{
  QUSA example for $K = 6$ (i.e. $N = 2^K = 64$ qubits representing the integers
  $z = \{0, 1, ..., 63\}$) with $u(z_* = 14) = 1$ and $u(z \neq 14) = 0$.
  Top panels: uniform superposition (left, Eq.~\ref{eq:initial}), followed by
  the application of $\mathcal{Q}_*$ (middle, Eq.~\ref{eq:q_star_psi_0}) and
  $\mathcal{Q}_+$ (right, Eq.~\ref{eq:psi_1}), where the amplitude of the state
  $\ket{14}$ is shown with a red dot and that of all other states with blue.
  Middle panels: the amplification of the amplitude of $\ket{14}$ after $2$
  (left), $3$ (middle), and $4$ (right) iterations (i.e. $2$, $3$, and $4$
  applications of the operator $\mathcal{Q}_+\mathcal{Q}_*$, respectively,
  see Eq.~\ref{eq:psi_m}).
  Bottom panel: amplitude amplification of $\ket{z_*}$ (red line,
  Eq.~\ref{eq:a_m_star}), the opposite effect for all other states (blue line,
  Eq.~\ref{eq:a_m}), and the optimal number of iterations (black dashed line,
  Eq.~\ref{eq:optimal_iterations}), which, for this example, is
  $\lceil(\pi\sqrt{N} - 2)/4\rceil = 6$.}
\label{fig:amplification}
\end{figure}

The initial uniform superposition of Eq.~(\ref{eq:initial}),
$\ket{\psi_0} = \ket{+}^{\otimes K}$, can be rewritten in terms of two
orthogonal states:
\begin{align}
\ket{\psi_0}
= \frac{1}{\sqrt{N}}\left(\sum_{z \neq z_*}\ket{z} + \ket{z_*}\right)
= \frac{\sqrt{N-1}}{\sqrt{N}}\ket{\zeta} + \frac{1}{\sqrt{N}}\ket{z_*}\,,
\end{align}
where $\ket{\zeta} \equiv (1/\sqrt{N-1})\sum_{z\neq z_*}\ket{z}$ and the
orthonormality of the basis states, $\braket{z_1}{z_2} = \delta_{z_1z_2}$,
implies that $\braket{\zeta}{\zeta} = 1$, $\braket{z_*}{z_*} = 1$, and
$\braket{\zeta}{z_*} = 0$; see top left panel of Fig.~\ref{fig:amplification}.
In the first iteration, the application of the operator
$\mathcal{Q}_+\mathcal{Q}_*$ gives:
\begin{align}
&\mathcal{Q}_*\ket{\psi_0}
= \Big(I - 2\ket{z_*}\bra{z_*}\Big)\ket{\psi_0}
= \ket{\psi_0} - \frac{2}{\sqrt{N}}\ket{z_*}\,, \label{eq:q_star_psi_0}\\
&\ket{\psi_1}
= \mathcal{Q}_+\mathcal{Q}_*\ket{\psi_0}
= \Big(2\ket{\psi_0}\bra{\psi_0} - I\Big)\mathcal{Q}_*\ket{\psi_0}
= \left(1-\frac{4}{N}\right)\frac{\sqrt{N-1}}{\sqrt{N}}\ket{\zeta}
  + \left(3-\frac{4}{N}\right)\frac{1}{\sqrt{N}}\ket{z_*}\,,
\label{eq:psi_1}
\end{align}
which amplifies the probability of measuring the state $\ket{z_*}$ by a factor
of $(3 - 4/N)^2$; see top middle and top right panels of
Fig.~\ref{fig:amplification}.
After $m$ iterations, we obtain the state
$\ket{\psi_m} = a_m\ket{\zeta} + a_{*m}\ket{z_*}$, where
$|a_m|^2 + |a_{*m}|^2 = 1$.
Iteration $m+1$ gives:
\begin{align}
\mathcal{Q}_*\ket{\psi_m}
&= \Big(I - 2\ket{z_*}\bra{z_*}\Big)\ket{\psi_m}
= \frac{a_m\sqrt{N}}{\sqrt{N-1}}\ket{\psi_0}
  - \frac{a_m + a_{*m}\sqrt{N-1}}{\sqrt{N-1}}\ket{z_*}\,, \nonumber\\
\ket{\psi_{m+1}}
&= \mathcal{Q}_+\mathcal{Q}_*\ket{\psi_m}
= a_{m+1}\ket{\zeta} + a_{*m+1}\ket{z_*}\,,
\label{eq:psi_m}
\end{align}
where
\begin{align}
a_{m+1} &= \frac{a_m(N - 2) - 2a_{*m}\sqrt{N-1}}{N}\,, \label{eq:a_m}\,\\
a_{*m+1} &= \frac{2a_m(N - 1) + a_{*m}(N - 2)\sqrt{N-1}}{N\sqrt{N-1}}\,,
\label{eq:a_m_star}
\end{align}
see the top and middle panels of Fig.~\ref{fig:amplification} for the first $4$
iterations.
Therefore, starting from the initial values $a_0 = \sqrt{N-1}/\sqrt{N}$ and
$a_{*0} = 1/\sqrt{N}$, we can calculate the amplitudes $a_m$ and $a_{*m}$ of the
$m$-th application of $\mathcal{Q}_+\mathcal{Q}_*$ recursively, see bottom panel
of Fig.~\ref{fig:amplification}.
To find how many iterations are needed, we first calculate the initial angle
between $\ket{z_*}$ and $\ket{\psi_0}$ which can be expressed as
$\pi/2 - \theta_\mathrm{ini}$, where $\theta_\mathrm{ini} \ll \pi/2$ is the
angle between $\ket{\zeta}$ and $\ket{\psi_0}$:
\begin{align}
\cos\left(\frac{\pi}{2} - \theta_\mathrm{ini}\right)
&= \frac{\braket{s}{z_*}}
  {\sqrt{\braket{s}{s}}\sqrt{\braket{z_*}{z_*}}} \quad\Rightarrow\quad
\sin\theta_\mathrm{ini} = \frac{1}{\sqrt{N}} \quad\Rightarrow\quad
\theta_\mathrm{ini} \simeq \frac{1}{\sqrt{N}}\,.
\end{align}
Then, note that every iteration rotates the vector $\ket{\psi_m}$ towards
$\ket{z_*}$ (giving the vector \ket{\psi_{m+1}}) by a constant angle
$\theta \simeq 2\theta_\mathrm{ini}$:
\begin{align}
\cos\theta &= \frac{\braket{\psi_{m+1}}{\psi_m}}
  {\sqrt{\braket{\psi}{\psi}}\sqrt{\braket{\psi_{m+1}}{\psi_{m+1}}}}
  \quad\Rightarrow\quad
1 - \frac{\theta^2}{2} \simeq a_{m+1}a_m + a_{*m+1}a_{*m} \quad\Rightarrow\quad
\theta \simeq \frac{2}{\sqrt{N}}\,.
\end{align}
The required number of iterations can be found by equating the initial angle
between $\mathrm{\psi_0}$ and $\ket{z_*}$, $\pi/2 - \theta_\mathrm{ini}$, with
$m$ rotations, $m\theta$, needed to align $\ket{\psi_m}$ with $\ket{z_*}$:
\begin{align}
m\theta &= \frac{\pi}{2} - \theta_\mathrm{ini} \quad\Rightarrow\quad
m \simeq \frac{\pi\sqrt{N} - 2}{4}\,,
\label{eq:optimal_iterations}
\end{align}
see bottom panel of Fig.~\ref{fig:amplification} (further iterations reduce the
probability of $\ket{z_*}$ because the amplitudes $a_m$ and $a_{*m}$ are
periodic functions).

\section{Operator $\mathcal{Q}_*$}

\subsection{Additional properties of the operators $\mathcal{B}_k$}
\label{app:commutation}

The application of $\mathcal{B}_k$ on a $K$-qubit register gives:
\begin{align}
&\mathcal{B}_k
  \ket{\mathrm{b}_{K-1}...\mathrm{b}_k...\mathrm{b}_1\mathrm{b}_0}
= \ket{\mathrm{b}_{K-1}}\otimes\cdots\otimes
  \mathcal{B}_k\ket{\mathrm{b}_k}
  \otimes\cdots\otimes
  \ket{\mathrm{b}_1}\otimes
  \ket{\mathrm{b}_0}
= x_k\ket{\mathrm{b}_{K-1}...\mathrm{b}_k...\mathrm{b}_1\mathrm{b}_0}\,.
\end{align}
The operators $\mathcal{B}_k$ commute with each other:
\begin{align}
[\mathcal{B}_{k_1}, \mathcal{B}_{k_2}]
  \ket{\mathrm{b}_{K-1}...\mathrm{b}_0}
&= \left(\mathcal{B}_{k_1}\mathcal{B}_{k_2}
    - \mathcal{B}_{k_2}\mathcal{B}_{k_1}\right)
  \ket{\mathrm{b}_{K-1}...\mathrm{b}_0}
= b_{k_2}\mathcal{B}_{k_1}
    \ket{\mathrm{b}_{K-1}...\mathrm{b}_0}
  - b_{k_1}\mathcal{B}_{k_2}
    \ket{\mathrm{b}_{K-1}...\mathrm{b}_0}
  \nonumber\\
&= \left(b_{k_2}b_{k_1} - b_{k_1}b_{k_2}\right)
  \ket{\mathrm{b}_{K-1}...\mathrm{b}_0}
= 0\ket{\mathrm{b}_{K-1}...\mathrm{b}_0}\,,
\end{align}
and also have the following useful property:
\begin{align}
\mathcal{B}_k^n\ket{\mathrm{b}_{K-1}...\mathrm{b}_1\mathrm{b}_0}
&= b_k^n\ket{\mathrm{b}_{K-1}...\mathrm{b}_1\mathrm{b}_0}
= b_k\ket{\mathrm{b}_{K-1}...\mathrm{b}_1\mathrm{b}_0}
= \mathcal{B}_k\ket{\mathrm{b}_{K-1}...\mathrm{b}_1\mathrm{b}_0} \,.
\end{align}

\subsection{Operator $\mathcal{Q}_*$ as reflection}
\label{app:reflection}

\begin{align}
Q_*\ket{z}
&= \lim_{n \to \infty} e^{i\pi\mathcal{G}^n}\ket{z}
= \lim_{n \to \infty}\left[
  I + i\pi\mathcal{G}^n + \frac{1}{2}i^2\pi^2\left(\mathcal{G}^n\right)^2
  + \cdots\right]\ket{z} \nonumber\\
&= \lim_{n \to \infty}\left[
  1 + i\pi g(\mathbf{x})^n + \frac{1}{2}i^2\pi^2\left[g(\mathbf{x})^n\right]^2
  + \cdots\right]\ket{z}
= \left[
  1 + i\pi u(z) + \frac{1}{2}i^2\pi^2u(z)^2
  + \cdots\right]\ket{z} \nonumber\\
&= \left[
  1 - u(z) + \left(1 + i\pi + \frac{1}{2}i^2\pi^2 + \cdots\right)u(z)
  \right]\ket{z}
= \left[1 - u(z) + e^{i\pi}u(z)\right]\ket{z} \nonumber\\
&= \Big[1 - 2u(z)\Big]\ket{z}
= \Big(I - 2\ket{z_*}\bra{z_*}\Big)\ket{z}
\,,
\end{align}
where we used $u(z)^n = u(z)$.
The last equation is implied from the fact that:
\begin{align}
\Big[1 - 2u(z)\Big]\ket{z \neq z_*}
&= 1\ket{z \neq z_*} - 0\ket{z \neq z_*}
= \ket{z \neq z_*}\,, \\
\Big[1 - 2u(z)\Big]\ket{z_*}
&= 1\ket{z_*} - 2\ket{z_*}
= -\ket{z_*}\,,
\end{align}
and
\begin{align}
\Big(I - 2\ket{z_*}\bra{z_*}\Big)\ket{z \neq z_*}
&= 1\ket{z \neq z_*} - 0\ket{z \neq z_*}
= \ket{z \neq z_*}\,, \\
\Big(I - 2\ket{z_*}\bra{z_*}\Big)\ket{z_*}
&= 1\ket{z_*} - 2\ket{z_*}
= -\ket{z_*}\,.
\end{align}

\subsection{Operator $\mathcal{Q}_{*0}$}
\label{app:Q0_operator}

\begin{align}
\mathcal{Q}_{*0}\ket{z}
&= e^{i\pi\mathcal{C}_0}
  \ket{\mathrm{b}_{K-1}...\mathrm{b}_1\mathrm{b}_0}
= \left(
  I + i\pi\mathcal{C}_0 + \frac{1}{2}i^2\pi^2\mathcal{C}_0^2
  + \frac{1}{3!}i^3\pi^3\mathcal{C}_0^3 + \cdots \right)
  \ket{\mathrm{b}_{K-1}...\mathrm{b}_1\mathrm{b}_0} \nonumber\\
&= \left(
  1 + i\pi c + \frac{1}{2}i^2\pi^2c^2 + \frac{1}{3!}i^3\pi^3c^3 + \cdots \right)
  \ket{\mathrm{b}_{K-1}...\mathrm{b}_1\mathrm{b}_0}
= e^{i\pi c}
  \ket{\mathrm{b}_{K-1}...\mathrm{b}_1\mathrm{b}_0} \nonumber\\
&= P(\phi_0)\ket{\mathrm{b}_{K-1}...\mathrm{b}_1\mathrm{b}_0}\,,
\end{align}
where $\phi_0 = c\pi$.

\subsection{Operator $\mathcal{Q}_{*1}$}
\label{app:Q1_operator}

\begin{align}
\mathcal{Q}_{*1}\ket{z}
&= e^{i\pi\mathcal{C}_1}
  \ket{\mathrm{b}_{K-1}...\mathrm{b}_1\mathrm{b}_0}
= e^{i\pi\sum_kc_k\mathcal{B}_k}
  \ket{\mathrm{b}_{K-1}...\mathrm{b}_1\mathrm{b}_0}
= \prod_{k=0}^{K-1} e^{i\pi c_k\mathcal{B}_k}
  \ket{\mathrm{b}_{K-1}...\mathrm{b}_1\mathrm{b}_0} \nonumber\\
&= \prod_k\left(
  I + i\phi_{1,k}\mathcal{B}_k + \frac{1}{2}i^2\phi_{1,k}^2\mathcal{B}_k^2
  + \frac{1}{3!}i^3\phi_{1,k}^3\mathcal{B}_k^3 + \cdots \right)
  \ket{\mathrm{b}_{K-1}...\mathrm{b}_1\mathrm{b}_0} \nonumber\\
&= \prod_k\left(
  I - \mathcal{B}_k + \mathcal{B}_k + i\phi_{1,k}\mathcal{B}_k
  + \frac{1}{2}i^2\phi_{1,k}^2\mathcal{B}_k + \cdots \right)
  \ket{\mathrm{b}_{K-1}...\mathrm{b}_1\mathrm{b}_0} \nonumber\\
&= \prod_k\left[
  I - \left(1 - e^{i\phi_{1,k}}\right)\mathcal{B}_k\right]
  \ket{\mathrm{b}_{K-1}...\mathrm{b}_1\mathrm{b}_0} \nonumber\\
&= \prod_k\mathcal{P}_{1,k}(\phi_{1,k})
  \ket{\mathrm{b}_{K-1}...\mathrm{b}_1\mathrm{b}_0}\,,
\end{align}
where $\phi_{1,k} = c_k\pi$, $\mathcal{P}_{1,k}$ is given by
\begin{align}
\mathcal{P}_{1,k}(\phi_{1,k})
&= \left\{\begin{array}{ll}
    I & \mathrm{if\ } b_k = 0 \\
    P(\phi_{1,k}) & \mathrm{if\ } b_k = 1
  \end{array}\right.\,,
\end{align}
and the proof for the last step is:
\begin{align}
\left[I - \left(1 - e^{i\phi}\right)\mathcal{B}\right]
  \left(a_0\ket{0} + a_1\ket{1}\right)
&= a_0\ket{0} + e^{i\phi}a_1\ket{1}
= P(\phi)\left(a_0\ket{0} + a_1\ket{1}\right)\,.
\end{align}

\subsection{Operator $\mathcal{Q}_{*2}$}
\label{app:Q2_operator}

\begin{align}
\mathcal{Q}_{*2}\ket{z}
&= e^{i\mathcal{C}_2}\ket{\mathrm{b}_{K-1}...\mathrm{b}_0}
= e^{i\pi\sum_{k_1}\sum_{k_2>k_1}c_{k_1k_2}
  \mathcal{B}_{k_2}\mathcal{B}_{k_1}}
  \ket{\mathrm{b}_{K-1}...\mathrm{b}_0} \nonumber\\
&= \prod_{k_1}\prod_{k_2>k_1}
    e^{i\phi_{2,k_1k_2}\mathcal{B}_{k_2}\mathcal{B}_{k_1}}
  \ket{\mathrm{b}_{K-1}...\mathrm{b}_0} \nonumber\\
&= \prod_{k_1}\prod_{k_2}\left(
  I + i\phi_{2,k_1k_2}\mathcal{B}_{k_2}\mathcal{B}_{k_1}
  + \frac{1}{2}i^2\phi_{2,k_1k_2}^2\mathcal{B}_{k_2}^2\mathcal{B}_{k_1}^2
  + \cdots \right)\ket{\mathrm{b}_{K-1}...\mathrm{b}_0} \nonumber\\
&= \prod_{k_1}\prod_{k_2}\left(
  I - \mathcal{B}_{k_2}\mathcal{B}_{k_1} + \mathcal{B}_{k_2}\mathcal{B}_{k_1}
  + i\phi_{2,k_1k_2}\mathcal{B}_{k_2}\mathcal{B}_{k_1}
  + \frac{1}{2}i^2\phi_{2,k_1k_2}^2\mathcal{B}_{k_2}\mathcal{B}_{k_1}
  + \cdots \right)\ket{\mathrm{b}_{K-1}...\mathrm{b}_0} \nonumber\\
&= \prod_{k_1}\prod_{k_2}\left[
  I - \left(1 - e^{i\phi_{2,k_1k_2}}\right)
    \mathcal{B}_{k_2}\mathcal{B}_{k_1}\right]
    \ket{\mathrm{b}_{K-1}...\mathrm{b}_0} \nonumber\\
&= \prod_{k_1}\prod_{k_2}\left[
  1 - \left(1 - e^{i\phi_{2,k_1k_2}}\right)
  b_{k_2}b_{k_1}\right]\ket{\mathrm{b}_{K-1}...\mathrm{b}_0} \nonumber\\
&= \prod_{k_1}\prod_{k_2}\mathcal{P}_{2,k_1k_2}(\phi_{2,k_1k_2})
  \ket{\mathrm{b}_{K-1}...\mathrm{b}_0}\,,
\end{align}
where $\phi_{2,k_1k_2} \equiv c_{k_1k_2}\pi$ and
\begin{align}
\mathcal{P}_{2,k_1k_2}(\phi_{2,k_1k_2})
&= \left\{\begin{array}{ll}
    I & \mathrm{if\ } b_{k_1}b_{k_2} = 0 \\
    P(\phi_{2,k_1k_2}) & \mathrm{if\ } b_{k_1}b_{k_2} = 1
  \end{array}\right.\,.
\end{align}

\subsection{Operator $\mathcal{Q}_{*j}$}
\label{app:Qn_operator}

\begin{align}
\mathcal{Q}_{*j}\ket{z}
&= e^{i\pi\mathcal{C}_j}
  \ket{\mathrm{b}_{K-1}...\mathrm{b}_0}
= \exp\left(i\pi\sum_{k_1}\cdots
  \!\!\!\!\!\!\!\!\sum_{k_j>k_1,...k_{j-1}}\!\!\!\!\!\!\!\!c_{k_1...k_j}
    \mathcal{B}_{k_j}\cdots\mathcal{B}_{k_1}\right)
  \ket{\mathrm{b}_{K-1}...\mathrm{b}_0} \nonumber\\
&= \prod_{k_j}\cdots\prod_{k_1} e^{i\phi_{j,k_1...k_j}
  \mathcal{B}_{k_j}\cdots\mathcal{B}_{k_1}}
  \ket{\mathrm{b}_{K-1}...\mathrm{b}_0} \nonumber\\
&= \prod_{k_j}\cdots\prod_{k_1}\left(
  I + i\phi_{j,k_1...k_j}\mathcal{B}_{k_j}\cdots\mathcal{B}_{k_1}
  + \frac{1}{2}i^2\phi_{j,k_1...k_j}^2
    \mathcal{B}_{k_j}^2\cdots\mathcal{B}_{k_1}^2
  + \cdots \right)\ket{\mathrm{b}_{K-1}...\mathrm{b}_0} \nonumber\\
&= \prod_{k_j}\cdots\prod_{k_1}\left(
  I - \mathcal{B}_{k_j}\cdots\mathcal{B}_{k_1}
  + \mathcal{B}_{k_j}\cdots\mathcal{B}_{k_1}
  + i\phi_{j,k_1...k_j}\mathcal{B}_{k_j}\cdots\mathcal{B}_{k_1}
  + \cdots \right)\ket{\mathrm{b}_{K-1}...\mathrm{b}_0} \nonumber\\
&= \prod_{k_j}\cdots\prod_{k_1}\left[
  I - \left(1 - e^{i\phi_{j,k_1...k_j}}\right)
  \mathcal{B}_{k_j}\cdots\mathcal{B}_{k_1}\right]
  \ket{\mathrm{b}_{K-1}...\mathrm{b}_0} \nonumber\\
&= \prod_{k_j}\cdots\prod_{k_1}\left[
  I - \left(1 - e^{i\phi_{j,k_1...k_j}}\right)
  \prod_{l=1}^jb_{k_l}\right]\ket{\mathrm{b}_{K-1}...\mathrm{b}_0} \nonumber\\
&= \prod_{k_j}\cdots\prod_{k_1}\mathcal{P}_{j,k_1...k_j}(\phi_{j,k_1...k_j})
  \ket{\mathrm{b}_{K-1}...\mathrm{b}_0}\,,
\end{align}
where $\phi_{j,k_1...k_j} \equiv c_{k_1...k_j}\pi$ and
\begin{align}
\mathcal{P}_{j,k_1...k_j}(\phi_{j,k_1...k_j})
&= \left\{\begin{array}{ll}
    I & \mathrm{if\ } \prod_{l=1}^jb_{k_l} = 0 \\
    P(\phi_{j,k_1...k_n}) & \mathrm{if\ } \prod_{l=1}^jb_{k_l} = 1
  \end{array}\right.\,.
\end{align}
The corresponding quantum gates for $k_1 < k_2 < \cdots < k_j$ is:
\begin{center}
\begin{quantikz}
\lstick{$\ket{b_0}\quad\,$}
  & \qw
    \gategroup[16,steps=9,style={dashed, rounded corners, inner xsep=2,
      inner ysep=16pt}]{$\mathcal{P}_{j,k_1...k_j}(\phi_{j,k_1...k_j})$}
    \gategroup[16,steps=4,style={dashed, rounded corners, inner xsep=0,
      inner ysep=0.2pt}]{$\mathrm{AND}$}
  & \qw
  & \cdots\qw
  & \qw
  & \qw
  & \qw\gategroup[16,steps=4,style={dashed, rounded corners, inner xsep=0,
    inner ysep=0.2pt}]{$\mathrm{AND}^\dagger$}
  & \cdots\qw
  & \qw
  & \qw
  & \qw \\
\wave&&&&&&&&&& \\
\lstick{$\ket{b_{k_1}}\,\,\,\,$}
  & \ctrl{2}
  & \qw
  & \cdots\qw
  & \qw
  & \qw
  & \qw
  & \cdots\qw
  & \qw
  & \ctrl{2}
  & \qw \\
\wave&&&&&&&&&&\\
\lstick{$\ket{b_{k_2}}\,\,\,\,$}
  & \ctrl{7}
  & \qw
  & \cdots\qw
  & \qw
  & \qw
  & \qw
  & \cdots\qw
  & \qw
  & \ctrl{7}
  & \qw \\
\wave&&&&&&&&&&\\
\lstick{$\ket{b_{k_3}}\,\,\,\,$}
  & \qw
  & \ctrl{5}
  & \cdots\qw
  & \qw
  & \qw
  & \qw
  & \cdots\qw
  & \ctrl{5}
  & \qw
  & \qw \\
\wave&&&&&&&&&&\\
\lstick{$\ket{b_{k_j}}\,\,\,\,$}
  & \qw
  & \qw
  & \cdots\qw
  & \ctrl{6}
  & \qw
  & \ctrl{6}
  & \cdots\qw
  & \qw
  & \qw
  & \qw \\
\wave&&&&&&&&&& \\
\lstick{$\ket{b_{K\!-\!1}}$}
  & \qw
  & \qw
  & \cdots\qw
  & \qw
  & \qw
  & \qw
  & \cdots\qw
  & \qw
  & \qw
  & \qw \\
\lstick{$\ket{0}_\textsc{and}^1$}
  & \targ{}
  & \ctrl{1}
  & \cdots\qw
  & \qw
  & \qw
  & \qw
  & \cdots\qw
  & \ctrl{1}
  & \targ{}
  & \qw \\
\lstick{$\ket{0}_\textsc{and}^2$}
  & \qw
  & \targ{}
  & \cdots\qw
  & \qw
  & \qw
  & \qw
  & \cdots\qw
  & \targ{}
  & \qw
  & \qw \\
\wave&&&&&&&&&& \\
\lstick{$\ket{0}_\textsc{and}^{\mathrm{j\!-\!1}}$}
  & \qw
  & \qw
  & \cdots\qw
  & \ctrl{1}
  & \qw
  & \ctrl{1}
  & \cdots\qw
  & \qw
  & \qw
  & \qw \\
\lstick{$\ket{0}_\mathrm{anc}\,$}
  & \qw
  & \qw
  & \cdots\qw
  & \targ{}
  & \gate{P(\phi_{j,k_1...k_j})}
  & \targ{}
  & \cdots\qw
  & \qw
  & \qw
  & \qw
\end{quantikz}
\end{center}
where $\mathrm{AND} = \mathrm{AND}^\dagger$ is an operator that flips the
ancilla qubit if $b_{k_1} = b_{k_2} = \cdots = b_{k_j} = 1$.

\section{The objective function example}
\label{app:f}

We can calculate the explicit form of Eq.~(\ref{eq:f}) as follows:
\begin{align}
f(\mathbf{x})
&= N_\mathrm{p}^2\left(\mu_\mathrm{p} - \mu_\mathrm{b}\right)^2
    + N_\mathrm{p}^4\left(\sigma_\mathrm{p}^2 - \sigma_\mathrm{b}^2\right)^2
    + \left(N_\mathrm{p} - N_\mathrm{b}\right)^2
    \nonumber\\
&= \left(\sum_ix_i\mu_i - \sum_ix_i\mu_\mathrm{b}\right)^2
  + \left(\sum_i\sum_jx_ix_j\sigma_i\sigma_j\rho_{ij}
    - \sum_i\sum_jx_ix_j\sigma_\mathrm{b}^2\right)^2
  + \left(\sum_ix_i - N_\mathrm{b}\right)^2 \nonumber\\
&= N_\mathrm{b}^2
  - 2\sum_ix_iN_\mathrm{b}
  + \sum_i\sum_jx_ix_j\left(
    1 + \mu_\mathrm{b}^2 - 2\mu_i\mu_\mathrm{b} + \mu_i\mu_j\right)
  \nonumber\\
  &\quad
  + \sum_i\sum_j\sum_k\sum_lx_ix_jx_kx_l\left(
    \sigma_\mathrm{b}^4
    - 2\sigma_i\sigma_j\rho_{ij}\sigma_\mathrm{b}^2
    + \sigma_i\sigma_j\sigma_k\sigma_l\rho_{ij}\rho_{kl}\right)\,.
\end{align}
For four assets, the function is:
\begin{align}
&f(x_\textsc{a},x_\textsc{b},x_\textsc{c},x_\textsc{d})
= w
  + w_\textsc{a}x_\textsc{a}
  + w_\textsc{b}x_\textsc{b}
  + w_\textsc{b}x_\textsc{c}
  + w_\textsc{b}x_\textsc{d} \nonumber\\
&\quad
  + w_\textsc{ab}x_\textsc{a}x_\textsc{b}
  + w_\textsc{ac}x_\textsc{a}x_\textsc{c}
  + w_\textsc{ad}x_\textsc{a}x_\textsc{d}
  + w_\textsc{bc}x_\textsc{b}x_\textsc{c}
  + w_\textsc{bd}x_\textsc{b}x_\textsc{d}
  + w_\textsc{cd}x_\textsc{c}x_\textsc{d} \nonumber\\
&\quad+
+ w_\textsc{abc}x_\textsc{a}x_\textsc{b}x_\textsc{c}
+ w_\textsc{abd}x_\textsc{a}x_\textsc{b}x_\textsc{d}
+ w_\textsc{acd}x_\textsc{a}x_\textsc{c}x_\textsc{d}
+ w_\textsc{bcd}x_\textsc{b}x_\textsc{c}x_\textsc{d}
+ w_\textsc{abcd}x_\textsc{a}x_\textsc{b}x_\textsc{c}x_\textsc{d}\,.
\end{align}
Similarly the functions $g(\mathbf{x})$ and $g(\mathbf{x})^n = u_n(z)$ are:
\begin{align}
g(\mathbf{x})
&= c_1 + c_{1,\textsc{a}}x_\textsc{a} + \cdots
  + c_{1,\textsc{abcd}}x_\textsc{a}x_\textsc{b}x_\textsc{c}x_\textsc{d}
\ \ \mathrm{and}
\ \ g(\mathbf{x})^n\!
= c_n + c_{n,\textsc{a}}x_\textsc{a} + \cdots
  + c_{n,\textsc{abcd}}x_\textsc{a}x_\textsc{b}x_\textsc{c}x_\textsc{d}.
\end{align}

\section{Quantum circuit}
\label{app:circuit}

\begin{center}
\begin{quantikz}[column sep=0.3cm]
\lstick{$\ket{0}_\textsc{a}\,\,\,\,\,$}
  & \gate{H}\qw
    \gategroup[7, steps=1, style={dashed, rounded corners, inner xsep=0,
      inner ysep=30}]{$\mathcal{Q}_H$}
  & \qw
    \gategroup[7, steps=3, style={dashed, rounded corners, inner xsep=0,
      inner ysep=15}]{$\mathcal{Q}_{*0}$}
    \gategroup[7, steps=15, style={dashed, rounded corners, inner xsep=2,
      inner ysep=30}]{$\mathcal{Q}_*$}
  & \qw
  & \qw
  & \ctrl{6}
    \gategroup[7, steps=3, style={dashed, rounded corners, inner xsep=0,
      inner ysep=0}]{$\mathcal{P}_{1,\textsc{a}}(\phi_{1,\textsc{a}})$}
    \gategroup[7, steps=12, style={dashed, rounded corners, inner xsep=0,
      inner ysep=15}]{$\mathcal{Q}_{*1}$}
  & \qw
  & \ctrl{6}
  & \qw
    \gategroup[7, steps=3, style={dashed, rounded corners, inner xsep=0,
      inner ysep=0}]{$\mathcal{P}_{1,\textsc{b}}(\phi_{1,\textsc{b}})$}
  & \qw
  & \qw
  & \qw
    \gategroup[7, steps=3, style={dashed, rounded corners, inner xsep=0,
      inner ysep=0}]{$\mathcal{P}_{1,\textsc{c}}(\phi_{1,\textsc{c}})$}
  & \qw
  & \qw
  & \qw
    \gategroup[7, steps=3, style={dashed, rounded corners, inner xsep=0,
      inner ysep=0}]{$\mathcal{P}_{1,\textsc{d}}(\phi_{1,\textsc{d}})$}
  & \qw
  & \qw
  & \cdots\qw \\
\lstick{$\ket{0}_\textsc{b}\,\,\,\,\,$}
  & \gate{H}\qw
  & \qw
  & \qw
  & \qw
  & \qw
  & \qw
  & \qw
  & \ctrl{5}
  & \qw
  & \ctrl{5}
  & \qw
  & \qw
  & \qw
  & \qw
  & \qw
  & \qw
  & \cdots\qw \\
\lstick{$\ket{0}_\textsc{c}\,\,\,\,\,$}
  & \gate{H}\qw
  & \qw
  & \qw
  & \qw
  & \qw
  & \qw
  & \qw
  & \qw
  & \qw
  & \qw
  & \ctrl{4}
  & \qw
  & \ctrl{4}
  & \qw
  & \qw
  & \qw
  & \cdots\qw \\
\lstick{$\ket{0}_\textsc{d}\,\,\,\,\,$}
  & \gate{H}\qw
  & \qw
  & \qw
  & \qw
  & \qw
  & \qw
  & \qw
  & \qw
  & \qw
  & \qw
  & \qw
  & \qw
  & \qw
  & \ctrl{3}
  & \qw
  & \ctrl{3}
  & \cdots\qw \\
\lstick{$\ket{0}_\textsc{and}^1$}
  & \qw
  & \qw
  & \qw
  & \qw
  & \qw
  & \qw
  & \qw
  & \qw
  & \qw
  & \qw
  & \qw
  & \qw
  & \qw
  & \qw
  & \qw
  & \qw
  & \cdots\qw \\
\lstick{$\ket{0}_\textsc{and}^2$}
  & \qw
  & \qw
  & \qw
  & \qw
  & \qw
  & \qw
  & \qw
  & \qw
  & \qw
  & \qw
  & \qw
  & \qw
  & \qw
  & \qw
  & \qw
  & \qw
  & \cdots\qw \\
\lstick{$\ket{0}_\mathrm{anc}\,$}
  & \qw
  & \gate{X}
  & \gate{P_{\phi_0}}
  & \gate{X}
  & \targ{}
  & \gate{P_{\phi_{1,\textsc{a}}}}
  & \targ{}
  & \targ{}
  & \gate{P_{\phi_{1,\textsc{b}}}}
  & \targ{}
  & \targ{}
  & \gate{P_{\phi_{1,\textsc{c}}}}
  & \targ{}
  & \targ{}
  & \gate{P_{\phi_{1,\textsc{d}}}}
  & \targ{}
  & \cdots\qw
\end{quantikz}
\end{center}
\begin{center}
\begin{quantikz}[row sep=0.3cm, column sep=0.25cm]
\cdots\qw
  & \ctrl{1}
    \gategroup[7, steps=3, style={dashed, rounded corners, inner xsep=0,
      inner ysep=0}]{$\mathcal{P}_{2,\textsc{ab}}(\phi_{2,\textsc{ab}})$}
    \gategroup[7, steps=18, style={dashed, rounded corners, inner xsep=0,
      inner ysep=15}]{$\mathcal{Q}_{*2}$}
    \gategroup[7, steps=18, style={dashed, rounded corners, inner xsep=2,
      inner ysep=30}]{$\mathcal{Q}_*$}
  & \qw
  & \ctrl{1}
  & \ctrl{2}
    \gategroup[7, steps=3, style={dashed, rounded corners, inner xsep=0,
      inner ysep=0}]{$\mathcal{P}_{2,\textsc{ac}}(\phi_{2,\textsc{ac}})$}
  & \qw
  & \ctrl{2}
  & \ctrl{3}
    \gategroup[7, steps=3, style={dashed, rounded corners, inner xsep=0,
      inner ysep=0}]{$\mathcal{P}_{2,\textsc{ad}}(\phi_{2,\textsc{ad}})$}
  & \qw
  & \ctrl{3}
  & \qw
    \gategroup[7, steps=3, style={dashed, rounded corners, inner xsep=0,
      inner ysep=0}]{$\mathcal{P}_{2,\textsc{bc}}(\phi_{2,\textsc{bc}})$}
  & \qw
  & \qw
  & \qw
    \gategroup[7, steps=3, style={dashed, rounded corners, inner xsep=0,
      inner ysep=0}]{$\mathcal{P}_{2,\textsc{bd}}(\phi_{2,\textsc{bd}})$}
  & \qw
  & \qw
  & \qw
    \gategroup[7, steps=3, style={dashed, rounded corners, inner xsep=0,
      inner ysep=0}]{$\mathcal{P}_{2,\textsc{cd}}(\phi_{2,\textsc{cd}})$}
  & \qw
  & \qw
  & \cdots\qw \\
\cdots\qw
  & \ctrl{5}
  & \qw
  & \ctrl{5}
  & \qw
  & \qw
  & \qw
  & \qw
  & \qw
  & \qw
  & \ctrl{1}
  & \qw
  & \ctrl{1}
  & \ctrl{2}
  & \qw
  & \ctrl{2}
  & \qw
  & \qw
  & \qw
  & \cdots\qw \\
\cdots\qw
  & \qw
  & \qw
  & \qw
  & \ctrl{4}
  & \qw
  & \ctrl{4}
  & \qw
  & \qw
  & \qw
  & \ctrl{4}
  & \qw
  & \ctrl{4}
  & \qw
  & \qw
  & \qw
  & \ctrl{1}
  & \qw
  & \ctrl{1}
  & \cdots\qw \\
\cdots\qw
  & \qw
  & \qw
  & \qw
  & \qw
  & \qw
  & \qw
  & \ctrl{3}
  & \qw
  & \ctrl{3}
  & \qw
  & \qw
  & \qw
  & \ctrl{3}
  & \qw
  & \ctrl{3}
  & \ctrl{3}
  & \qw
  & \ctrl{3}
  & \cdots\qw \\
\cdots\qw
  & \qw
  & \qw
  & \qw
  & \qw
  & \qw
  & \qw
  & \qw
  & \qw
  & \qw
  & \qw
  & \qw
  & \qw
  & \qw
  & \qw
  & \qw
  & \qw
  & \qw
  & \qw
  & \cdots\qw \\
\cdots\qw
  & \qw
  & \qw
  & \qw
  & \qw
  & \qw
  & \qw
  & \qw
  & \qw
  & \qw
  & \qw
  & \qw
  & \qw
  & \qw
  & \qw
  & \qw
  & \qw
  & \qw
  & \qw
  & \cdots\qw \\
\cdots\qw
  & \targ{}
  & \gate{P_{\phi_{2,\textsc{ab}}}}
  & \targ{}
  & \targ{}
  & \gate{P_{\phi_{2,\textsc{ac}}}}
  & \targ{}
  & \targ{}
  & \gate{P_{\phi_{2,\textsc{ad}}}}
  & \targ{}
  & \targ{}
  & \gate{P_{\phi_{2,\textsc{bc}}}}
  & \targ{}
  & \targ{}
  & \gate{P_{\phi_{2,\textsc{bd}}}}
  & \targ{}
  & \targ{}
  & \gate{P_{\phi_{2,\textsc{cd}}}}
  & \targ{}
  & \cdots\qw
\end{quantikz}
\end{center}
\begin{center}
\begin{quantikz}[row sep=0.3cm, column sep=0.25cm]
\cdots\qw
  & \ctrl{1}
    \gategroup[7, steps=5, style={dashed, rounded corners, inner xsep=0,
      inner ysep=0}]{$\mathcal{P}_{3,\textsc{abc}}(\phi_{3,\textsc{abc}})$}
    \gategroup[7, steps=20, style={dashed, rounded corners, inner xsep=0,
      inner ysep=15}]{$\mathcal{Q}_{*3}$}
    \gategroup[7, steps=20, style={dashed, rounded corners, inner xsep=2,
      inner ysep=30}]{$\mathcal{Q}_*$}
  & \qw
  & \qw
  & \qw
  & \ctrl{1}
  & \ctrl{1}
    \gategroup[7, steps=5, style={dashed, rounded corners, inner xsep=0,
      inner ysep=0}]{$\mathcal{P}_{3,\textsc{abd}}(\phi_{3,\textsc{abd}})$}
  & \qw
  & \qw
  & \qw
  & \ctrl{1}
  & \ctrl{3}
    \gategroup[7, steps=5, style={dashed, rounded corners, inner xsep=0,
      inner ysep=0}]{$\mathcal{P}_{3,\textsc{acd}}(\phi_{3,\textsc{acd}})$}
  & \qw
  & \qw
  & \qw
  & \ctrl{3}
  & \qw
    \gategroup[7, steps=5, style={dashed, rounded corners, inner xsep=0,
      inner ysep=0}]{$\mathcal{P}_{3,\textsc{bcd}}(\phi_{3,\textsc{bcd}})$}
  & \qw
  & \qw
  & \qw
  & \qw
  & \cdots\qw \\
\cdots\qw
  & \ctrl{3}
  & \qw
  & \qw
  & \qw
  & \ctrl{3}
  & \ctrl{3}
  & \qw
  & \qw
  & \qw
  & \ctrl{3}
  & \qw
  & \qw
  & \qw
  & \qw
  & \qw
  & \ctrl{1}
  & \qw
  & \qw
  & \qw
  & \ctrl{1}
  & \cdots\qw \\
\cdots\qw
  & \qw
  & \ctrl{2}
  & \qw
  & \ctrl{2}
  & \qw
  & \qw
  & \qw
  & \qw
  & \qw
  & \qw
  & \ctrl{2}
  & \qw
  & \qw
  & \qw
  & \ctrl{2}
  & \ctrl{2}
  & \qw
  & \qw
  & \qw
  & \ctrl{2}
  & \cdots\qw \\
\cdots\qw
  & \qw
  & \qw
  & \qw
  & \qw
  & \qw
  & \qw
  & \ctrl{1}
  & \qw
  & \ctrl{1}
  & \qw
  & \qw
  & \ctrl{1}
  & \qw
  & \ctrl{1}
  & \qw
  & \qw
  & \ctrl{1}
  & \qw
  & \ctrl{1}
  & \qw
  & \cdots\qw \\
\cdots\qw
  & \targ{}
  & \ctrl{2}
  & \qw
  & \ctrl{2}
  & \targ{}
  & \targ{}
  & \ctrl{2}
  & \qw
  & \ctrl{2}
  & \targ{}
  & \targ{}
  & \ctrl{2}
  & \qw
  & \ctrl{2}
  & \targ{}
  & \targ{}
  & \ctrl{2}
  & \qw
  & \ctrl{2}
  & \targ{}
  & \cdots\qw \\
\cdots\qw
  & \qw
  & \qw
  & \qw
  & \qw
  & \qw
  & \qw
  & \qw
  & \qw
  & \qw
  & \qw
  & \qw
  & \qw
  & \qw
  & \qw
  & \qw
  & \qw
  & \qw
  & \qw
  & \qw
  & \qw
  & \cdots\qw \\
\cdots\qw
  & \qw
  & \targ{}
  & \gate{P_{\phi_{3,\textsc{abc}}}}
  & \targ{}
  & \qw
  & \qw
  & \targ{}
  & \gate{P_{\phi_{3,\textsc{abd}}}}
  & \targ{}
  & \qw
  & \qw
  & \targ{}
  & \gate{P_{\phi_{3,\textsc{acd}}}}
  & \targ{}
  & \qw
  & \qw
  & \targ{}
  & \gate{P_{\phi_{3,\textsc{bcd}}}}
  & \targ{}
  & \qw
  & \cdots\qw
\end{quantikz}
\end{center}
\begin{center}
\begin{quantikz}[row sep=0.3cm, column sep=0.28cm]
\cdots\qw
  & \ctrl{1}
    \gategroup[7, steps=7, style={dashed, rounded corners, inner xsep=0,
      inner ysep=0}]{$\mathcal{P}_{4,\textsc{abcd}}(\phi_{4,\textsc{abcd}})$}
    \gategroup[7, steps=7, style={dashed, rounded corners, inner xsep=0,
      inner ysep=15}]{$\mathcal{Q}_{*4}$}
    \gategroup[7, steps=7, style={dashed, rounded corners, inner xsep=2,
      inner ysep=30}]{$\mathcal{Q}_*$}
  & \qw
  & \qw
  & \qw
  & \qw
  & \qw
  & \ctrl{1}
  & \gate{H}
    \gategroup[7, steps=1, style={dashed, rounded corners, inner xsep=2,
      inner ysep=15}]{$\mathcal{Q}_H$}
    \gategroup[7, steps=13, style={dashed, rounded corners, inner xsep=2,
      inner ysep=30}]{$\mathcal{Q}_+$}
  & \gate{X}
    \gategroup[7, steps=11, style={dashed, rounded corners, inner xsep=2,
      inner ysep=15}]{$\mathcal{Q}_0$}
  & \ctrl{1}
  & \qw
  & \qw
  & \qw
  & \qw
  & \qw
  & \qw
  & \qw
  & \ctrl{1}
  & \gate{X}
  & \gate{H}
    \gategroup[7, steps=1, style={dashed, rounded corners, inner xsep=2,
      inner ysep=15}]{$\mathcal{Q}_H^\dagger$}
  & \cdots\qw \\
\cdots\qw
  & \ctrl{3}
  & \qw
  & \qw
  & \qw
  & \qw
  & \qw
  & \ctrl{3}
  & \gate{H}
  & \gate{X}
  & \ctrl{3}
  & \qw
  & \qw
  & \qw
  & \qw
  & \qw
  & \qw
  & \qw
  & \ctrl{3}
  & \gate{X}
  & \gate{H}
  & \cdots\qw \\
\cdots\qw
  & \qw
  & \ctrl{2}
  & \qw
  & \qw
  & \qw
  & \ctrl{2}
  & \qw
  & \gate{H}
  & \gate{X}
  & \qw
  & \ctrl{2}
  & \qw
  & \qw
  & \qw
  & \qw
  & \qw
  & \ctrl{2}
  & \qw
  & \gate{X}
  & \gate{H}
  & \cdots\qw \\
\cdots\qw
  & \qw
  & \qw
  & \ctrl{2}
  & \qw
  & \ctrl{2}
  & \qw
  & \qw
  & \gate{H}
  & \gate{X}
  & \qw
  & \qw
  & \ctrl{2}
  & \qw
  & \qw
  & \qw
  & \ctrl{2}
  & \qw
  & \qw
  & \gate{X}
  & \gate{H}
  & \cdots\qw \\
\cdots\qw
  & \targ{}
  & \ctrl{1}
  & \qw
  & \qw
  & \qw
  & \ctrl{1}
  & \targ{}
  & \qw
  & \qw
  & \targ{}
  & \ctrl{1}
  & \qw
  & \qw
  & \qw
  & \qw
  & \qw
  & \ctrl{1}
  & \targ{}
  & \qw
  & \qw
  & \cdots\qw \\
\cdots\qw
  & \qw
  & \targ{}
  & \ctrl{1}
  & \qw
  & \ctrl{1}
  & \targ{}
  & \qw
  & \qw
  & \qw
  & \qw
  & \targ{}
  & \ctrl{1}
  & \qw
  & \qw
  & \qw
  & \ctrl{1}
  & \targ{}
  & \qw
  & \qw
  & \qw
  & \cdots\qw \\
\cdots\qw
  & \qw
  & \qw
  & \targ{}
  & \gate{P_{\phi_{4,\textsc{abcd}}}}
  & \targ{}
  & \qw
  & \qw
  & \qw
  & \qw
  & \qw
  & \qw
  & \targ{}
  & \gate{X}
  & \gate{Z}
  & \gate{X}
  & \targ{}
  & \qw
  & \qw
  & \qw
  & \qw
  & \cdots\qw
\end{quantikz}
\end{center}

\end{document}